\documentclass[a4paper,11pt]{article}
\author{{\Large Sibylle Driezen}\vspace{.6cm}\\
Instituto Galego de F\'isica de Altas Enerx\'ias (IGFAE),\\ Universidade de Santiago de Compostela, Spain \vspace{.6cm}\\  \href{mailto:sib.driezen@gmail.com}{\texttt{sib.driezen@gmail.com}}     } 
\date{}
\title{Modave Lectures on Classical Integrability in $2d$ Field Theories}

\usepackage{mathtools}
\usepackage{amsthm}
\usepackage{amsmath,amssymb,euscript,array,mathrsfs,appendix,marvosym,bbm,enumitem}
\usepackage{arydshln}
\usepackage{todonotes}
\usepackage{graphicx}
\usepackage{float}
\usepackage[margin=1cm,font=small]{caption}
\usepackage[margin=.2cm,font=footnotesize]{subcaption}
\usepackage[normalem]{ulem}
\usepackage{mdframed}
\usepackage{bm}
\usepackage{cite}

\newtheorem{theorem}{Theorem}[section]
\newtheorem{exercise}[theorem]{Exercise}

\usepackage[linkcolor=blue,colorlinks=true]{hyperref}
\usepackage{cleveref}
\hypersetup{
    colorlinks,
    linkcolor={blue},
    citecolor={blue},
    urlcolor={purple}
}

\usepackage[bottom]{footmisc}

\usepackage{wasysym}
\usepackage{enumitem}

 \usepackage{xcolor}

\usepackage{geometry}
\renewenvironment{proof}{\noindent{\bfseries Proof.}}{\qed \newline}
 
 \mdfdefinestyle{MyFrame}{%
    linecolor=black,
    linewidth=1pt,
    roundcorner=10pt,
    innertopmargin=\baselineskip,
    innerbottommargin=\baselineskip,
    innerrightmargin=20pt,
    innerleftmargin=20pt,
    userdefinedwidth=40em,
    align=center
    }

\usepackage{enumitem}

\newcommand{\Tr}{\mathrm{Tr}}

\DeclareFontFamily{U}{mathx}{\hyphenchar\font45}
\DeclareFontShape{U}{mathx}{m}{n}{<-> mathx10}{}
\DeclareSymbolFont{mathx}{U}{mathx}{m}{n}
\DeclareMathAccent{\widebar}{0}{mathx}{"73}

\DeclareMathAlphabet{\mathdsl}{U}{bbm}{m}{sl}
 
\makeatletter
\DeclareFontFamily{OMX}{MnSymbolE}{}
\DeclareSymbolFont{MnLargeSymbols}{OMX}{MnSymbolE}{m}{n}
\SetSymbolFont{MnLargeSymbols}{bold}{OMX}{MnSymbolE}{b}{n}
\DeclareFontShape{OMX}{MnSymbolE}{m}{n}{
    <-6>  MnSymbolE5
   <6-7>  MnSymbolE6
   <7-8>  MnSymbolE7
   <8-9>  MnSymbolE8
   <9-10> MnSymbolE9
  <10-12> MnSymbolE10
  <12->   MnSymbolE12
}{}
\DeclareFontShape{OMX}{MnSymbolE}{b}{n}{
    <-6>  MnSymbolE-Bold5
   <6-7>  MnSymbolE-Bold6
   <7-8>  MnSymbolE-Bold7
   <8-9>  MnSymbolE-Bold8
   <9-10> MnSymbolE-Bold9
  <10-12> MnSymbolE-Bold10
  <12->   MnSymbolE-Bold12
}{}

\let\llangle\@undefined
\let\rrangle\@undefined
\DeclareMathDelimiter{\llangle}{\mathopen}%
                     {MnLargeSymbols}{'164}{MnLargeSymbols}{'164}
\DeclareMathDelimiter{\rrangle}{\mathclose}%
                     {MnLargeSymbols}{'171}{MnLargeSymbols}{'171}               
\makeatother
\usepackage{stmaryrd}
 
 \numberwithin{equation}{section}

\begin{document}
\maketitle
\begin{abstract}
These lecture notes are based on a blackboard course given at the XVII Modave Summer School in Mathematical Physics held from 13---17 September 2021 in Brussels (Belgium), and aimed at Ph.D.~students in High Energy Theoretical Physics.
We start with introducing classical integrability in finite-dimensional systems to set the stage for our main purpose: introducing two-dimensional classical field theories which are integrable. We focus on their zero-curvature formulation through the  so-called Lax connection, which ensures the existence of an infinite tower of conserved charges. We then move on to their Poisson bracket structure, known as the Sklyanin or Maillet structure,  which ensures complete classical integrability. All the concepts that we encounter will be illustrated with the integrable Principal Chiral Model, which is the canonical sigma-model  that appears (or its generalisations) on the worldsheet of many string backgrounds.
Along the way we briefly comment on the properties of  integrability at the quantum level.
\end{abstract}

\newpage

\tableofcontents

\vspace{.5cm}

%----------------------------------------------------------------------------------------

% Define some commands to keep the formatting separated from the content 
%\newcommand{\keyword}[1]{\textbf{#1}}
%\newcommand{\tabhead}[1]{\textbf{#1}}
%\newcommand{\code}[1]{\texttt{#1}}
%\newcommand{\file}[1]{\texttt{\bfseries#1}}
%\newcommand{\option}[1]{\texttt{\itshape#1}}

%----------------------------------------------------------------------------------------

\newpage

\section{Introduction}

These lecture notes serve as an introduction to the vast subject of classical integrability in two-dimensional field theories and  sigma models. Integrable models are particularly appealing because they possess  a large number of conserved charges. Whilst the dynamics can be very complicated  and non-linear, the property of integrability allows to make progress by  providing a large toolkit of mathematical methods that often renders these theories exactly solvable. Let us  stress, however, that having the possibility of using  integrable methods  does not necessarily mean that one can in fact solve the system. Indeed, instead of \textit{solvability} one should consider having \textit{exact methods} as the main merit of an integrable model. At the classical level, well-known methods  are Liouville's quadratures, the Classical Inverse Scattering Method and the Classical Spectral Curve. At the quantum level, the main ones are the Quantum Inverse Scattering Method, Bethe Ansatz Techniques and the Quantum Spectral Curve.

Integrable models appear in various fields of mathematical physics with a ranging number of applications. 
They arise in fluid mechanics (e.g.~the Korteweg-De Vries model) and several areas of condensed matter physics (e.g.~the Sine-Gordon model).
They provide two-dimensional toy models for four-dimensional strongly coupled gauge theories (e.g.~the Principal Chiral Model) including phenomena such as asymptotic freedom and the  dynamical generation of a mass gap in the IR, which can be described non-perturbatively because of integrability.
 They also have a close connection to mathematical algebra theory. In particular,  integrable  structures of physical models have provided  physical realisations (through e.g.~the existence of hidden symmetries)  of algebraic structures such as Yangian algebras and affine quantum groups. Although we are not exhaustive, let us finally mention that integrable models also arise on the worldsheet of many instances of string backgrounds relevant for the AdS/CFT duality, such as $\textrm{AdS}_5\times \textrm{S}^5$. 	For the string theorist reading these notes this is perhaps the most interesting application. Indeed, together with the appearance of integrability in $N=4$ Super-Yang-Mills (SYM), this property has  resulted in remarkable and rigorous tests of AdS/CFT in the planar limit. In particular,  the exact spectrum of anomalous dimensions was computed and was found to interpolate between perturbative results on both sides of the duality. \\
 
These notes are organised as follows. In chapter \ref{s:Finite}, we first introduce classical integrability for finite-dimensional systems. After a short recap on the canonical formulation in section \ref{s:can}  we define   integrability in terms of Liouville's formulation in section \ref{s:Liouville}.  We show how the method of Liouville's quadratures can be used to solve the equations of motion in a very general way by  transforming to a  simple set of phase space variables which have solutions linear in time. Another convenient set of variables,  the action-angle variables, which are adapted to the global dynamics of an integrable system are defined in \ref{s:AA}.
In  section \ref{s:LaxPairFinite} we then introduce the  Lax pair formulation, which is more convenient to generalise to integrable field theories. 
 We illustrate all the concepts of chapter \ref{s:Finite} with the simple example of an anisotropic harmonic oscillator in section \ref{s:exampleHO} and \ref{s:exampleHO2}.
  Building on the Lax pair formulation, we  introduce classical integrability for two-dimensional field theories in chapter \ref{s:IFT}. 
We define the (weak) classical integrable structure in section \ref{s:ClassicalIFT} through the zero-curvature formulation, which requires the existence of a  Lax connection whose flatness represents the equations of motion. We then show how this property ensures the existence of an infinite tower of conserved charges generated by the so-called monodromy matrix. In section \ref{s:ClassMethods}, we comment on how the weak integrable structure opens the possibility of using integrable methods such as the Classical Inverse Scattering  and the Classical Spectral Curve. The strong version of integrability is  defined in section \ref{s:FTInvolution} by requiring a particular structure of the Poisson brackets,  known as the  Sklyanin or Maillet form, through the existence of an $r/s$-matrix. We end this chapter by commenting on the main hallmarks of the quantum integrable regime and how these are influenced by the algebraic properties of $r$-matrices in section \ref{s:qint}. In chapter \ref{s:PCM} we illustrate all these concepts for the integrable Principal Chiral Model (PCM), which is a non-linear sigma model relevant for string worldsheet theories. For this reason we first briefly recap the main features of non-linear string sigma models in section \ref{s:sigmamodels}. The PCM model itself is then introduced in section \ref{s:PCMaction}, and its weak and strong classical integrable structure  is  discussed in sections \ref{s:PCMint} and \ref{s:PCMMaillet}. The construction of several infinite towers of conserved charges is given in quite some detail. In section \ref{s:intdef} we finally give some brief comments on  integrable deformations of sigma models. We end with a short conclusion in chapter \ref{s:Conclusions} in which we also comment on some of the possible directions in the broad area of integrability that we did not take. \\

\noindent \textit{Literature.}  As we mentioned, there will be  several subjects that we will not treat in these lecture notes. Therefore, let us refer to some excellent books, notes and reviews that we followed, and which the reader can consult when their interest is triggered. Within these notes they will also be mentioned at appropriate places. Non-exhaustively, we recommend the books \cite{Babelon:2003qtg} for classical integrability in general, and \cite{Faddeev:1987ph} for classical integrability of sigma models. There are also several excellent lecture notes and reviews: \cite{Torrielli:2016ufi} on classical integrability, \cite{Loebbert:2016cdm} on Yangian symmetry, \cite{Zarembo:2017muf} on integrable sigma models, \cite{Hoare:LN} on integrable \textit{string} sigma models, and  \cite{Beisert:2010jr,Arutyunov:2009ga} on integrability in AdS/CFT. A very nice collection of lecture notes ranging from classical integrability (e.g.~classical inverse scattering method, symmetries, \ldots) to quantum integrability (symmetries, S-matrices and Bethe Ansatz techniques) is also given by\cite{Bombardelli:2016rwb}. Recent notes are \cite{Retore:2021wwh} on classical and quantum integrability and \cite{Hoare:2021dix} on integrable deformations of sigma models.. Finally, we recommend also the notes \cite{Thompson:2019ipl} which shows the close connection between integrable deformations of sigma models and generalised worldsheet dualities (i.e.~non-abelian and Poisson-Lie T-duality) which has attracted a lot of interest in recent years.

\newpage
\section{Classical integrability in finite-dimensional  systems} \label{s:Finite}

\subsection{Recap on the canonical formulation} \label{s:can}

For finite-dimensional classical systems the notion of classical integrability has a straightforward definition given in the canonical formulation, and which is based on Liouville's theorem. Let us therefore first recall some generic facts about the canonical formulation of classical systems. 

We denote the number degrees of freedom by $n$, the coordinates by $q_i$, with $i \in \{ 1, \ldots, n\}$ and the conjugate momenta by $p_i$.  
 The state of the classical system is a point $(q_i, p_i)$ in the phase space $M$ which is a symplectic manifold carrying a (non-degenerate)  Poisson structure  $\{ \cdot , \cdot \}$. For  any $F,G \in {\cal F}(M)$, i.e.~functions on phase space, the Poisson bracket is defined as
\begin{equation}
\{ F, G \} = \frac{\partial F}{\partial p_i} \frac{\partial G}{\partial q_i} - \frac{\partial F}{\partial q_i} \frac{\partial G}{\partial p_i} .
\end{equation}
It is antisymmetric, satisfies the Leibniz rule and the Jacobi identity.\footnote{These properties imply that the non-degenerate Poisson bracket has a one-to-one correspondence with a symplectic form $\omega$, i.e.~a non-degenerate closed two-form.} The  Poisson bracket thus defines a Lie algebra on $ {\cal F}(M)$. 
In particular, to any function $F \in {\cal F}(M)$ we can associate a vector field denoted by $X_F$ as
\begin{equation} \label{eq:MfVectors}
X_F = \{ F, \cdot \} =  \frac{\partial F}{\partial q_i} \frac{\partial }{\partial p_i} - \frac{\partial F}{\partial p_i} \frac{\partial }{\partial q_i} .
\end{equation}
The map $F \mapsto X_F$ defines a homomorphism of Lie algebras $[X_F , X_G ] = X_{\{F,G\}}$. \newline

\begin{exercise}
Show  $[X_F , X_G ] = X_{\{F,G\}}$.
\end{exercise}

\vspace{.2cm}

\begin{figure}[h]
\centering
\includegraphics[scale=.36]{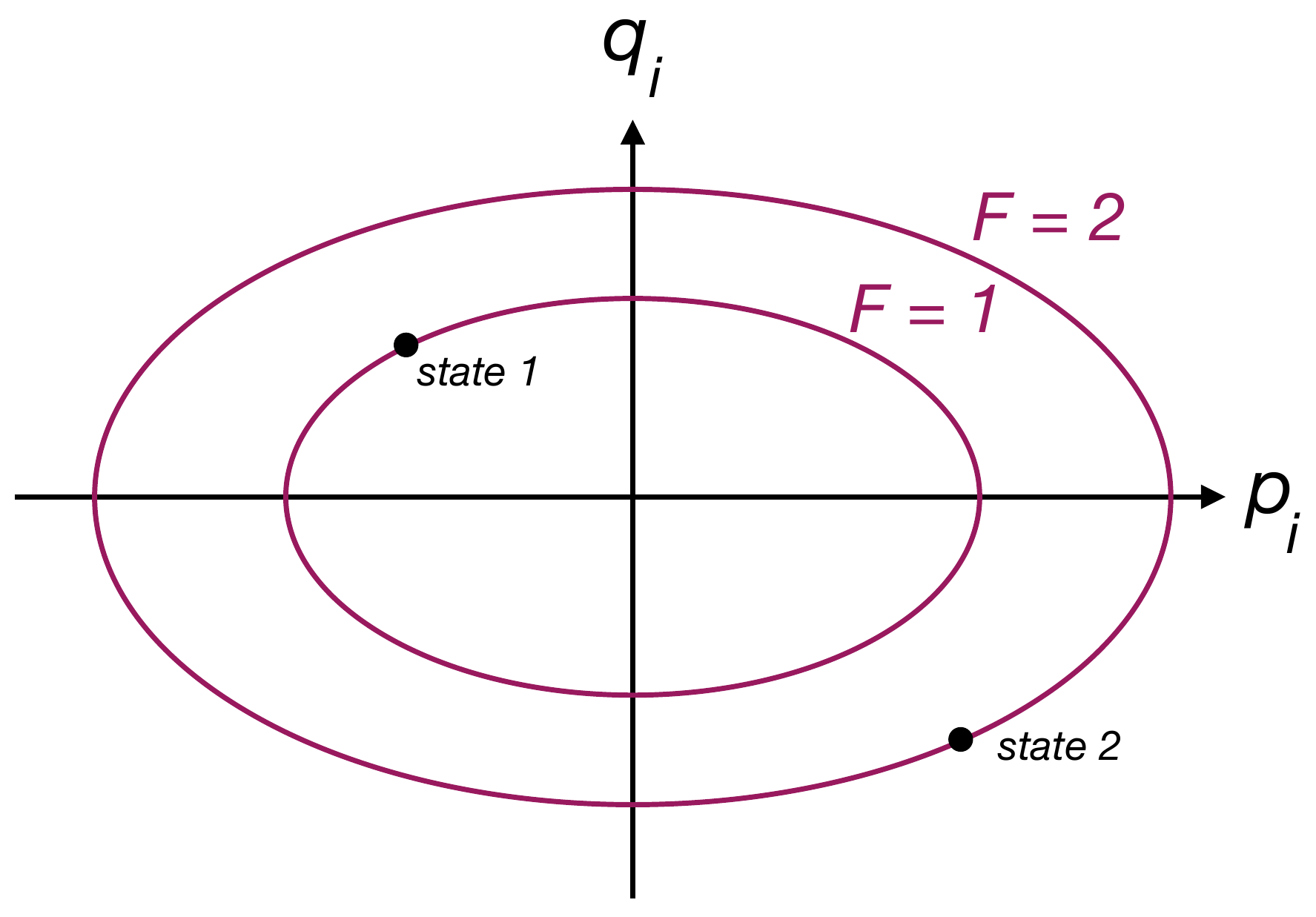}
\caption{Illustration of the phase space of a classical system with a conserved quantity $F$, such that  motion is restricted to the purple submanifolds.}
\label{f:submanifolds}
\end{figure}

\noindent Time $\tau$ evolution of the states of the system, or more generally functions on phase space, are generated by the Hamiltonian $H$, which itself is a function $H(q_i, p_i)$ on $M$.
 First recall that in this formulation the equations of motion (EOMs) of states are a system of $2n$ first order differential equations given by,
\begin{equation}
\dot{q_i} = \frac{\partial H}{\partial p_i} , \qquad \dot{p_i} = - \frac{\partial H}{\partial q_i} ,
\end{equation}
where the dot refers to time derivation. This implies that for any function $F\in {\cal F}(M)$ of phase space we have that,
\begin{equation}
\dot F = \{H , F \}  \ .
\end{equation}
When  $\{H, F\}= X_H (F) =  0$, then $F(\text{state}) = f$ is a constant in time, and thus the function $F$ is a conserved quantity. The trajectory of the state is then restricted to a submanifold of phase space, denoted by $M_f$, on which $F$ remains constant.  These surfaces can not intersect each other, and thus must foliate phase space. See figure \ref{f:submanifolds} for an illustration. Notice that  $\dot{H} = 0$ automatically. 

 When describing the dynamics of a specific classical system, what we always try to do is to find  solutions of its equation of motion. In  full generality, it is difficult (if not impossible) to solve this coupled partial differential system explicitly. When we have a conserved quantity, however, we see from figure \ref{f:submanifolds}  that the dynamics of phase space already exhibits a certain structure.  When the system is furthermore Liouville integrable, the problem of solving the EOMs simplifies  significantly.  We will see that in this case only a finite number of \textit{algebraic} equations need to be solved, rather than  differential equations, and thus  the system becomes exactly solvable. Let us now not postpone  any further and define  Liouville integrability.

\subsection{Liouville integrability} \label{s:Liouville}
A classical dynamical system is \textit{Liouville integrable} when 
\begin{itemize}
\item it has $n$   \textit{conserved quantities} $F_i$, $i  \in \{ 1, \cdots , n\}$, i.e.~as much as there are degrees of freedom,
\begin{equation}
\dot{ F}_i = 0 \quad \Leftrightarrow \quad \{ H, F_i \} = 0,
\end{equation}
which are globally defined on $M$, and
\item these conserved quantities are all \textit{in involution}, i.e.~they all Poisson commute amongst each other (in addition to commuting with the Hamiltonian only)
\begin{equation}
\{ F_i , F_j \} = 0 ,
\end{equation}
for all $i,j \in \{1, \cdots , n\}$, and 
\item these conserved quantities are all \textit{independent}, which refers to the linear independence of the following one-forms on $M$
\begin{equation}
dF_i = \frac{\partial F_i}{\partial q_j} dq_j + \frac{\partial F_i}{\partial p_j} dp_j \ .
\end{equation}
This is equivalent to having, at generic points, linearly independent  gradient vectors of $F_i$.
\end{itemize}

\noindent The above three properties imply that the \textit{level set} $M_f$ defined by the constants of motion $F_i ( p,q) = f_i$ as
\begin{equation}
M_f = \{ x = (q_i,p_i) \in M \ : \ F_i (x)= f_i \in \mathbb{R}, \  i = 1, \ldots , n \} ,
\end{equation}
has an $n$-dimensional tangent space at each point in $M_f$. A good basis for the tangent space is given by the vectors $X_{F_i}$ defined in \eqref{eq:MfVectors}, because  they are in fact linearly independent and tangent to $M_f$. The latter follows from the involution property
\begin{equation}
X_{F_i} (F_j) = \{ F_j , F_i \} = 0 \ .
\end{equation}
%
%
% Having $n$ \textit{independent} conserved quantities $F_i$ means that at generic points their gradient vectors, or equivalently the one-forms $d F_i = \frac{\partial F_i}{\partial q_j} dq_j + \frac{\partial F_i}{\partial p_j} dp_j$, must be linearly independent. 
%In other words, at generic points the tangent space of $M_f$ exists and is $n$-dimensional. 
%This implies that the vector fields $X_{F_i}$ are tangent to $M_f$ and form a consistent (linearly independent) basis for the tangent space at each point of $M_f$
Thus  for a generic state, the level set $M_f$ is an $n$-dimensional \textit{submanifold} which foliates phase space  globally. Let us finally remark that the maximal number of independent conserved quantities in involution is $n$. This implies that the Hamiltonian $H(q_i , p_i )$ can be rewritten as a function of the conserved quantities $ H(F_i)$. \newline

\begin{exercise}
Proof that there can not be more than $n$ independent conserved quantities in involution. 
\end{exercise}

\vspace{.2cm}

\noindent {\bf Liouville's theorem} ---
The theorem of Liouville states that in a Liouville integrable system one can perform a canonical transformation from the canonical coordinates $(p_i , q_i)$ to one where the new conjugate momenta coincide with the conserved quantities $F_i$, in other words $(p_i , q_i) \rightarrow (p'_i , q'_i)=(F_i , \psi_i)$ for some $\psi_i$. In this new basis, the equations of motion are decoupled and have  solutions linear in time. In principle this means that, in the canonical basis, 
 the equations of motion of Liouville integrable systems  can be fully solved  by performing a finite number of integrals and solving a finite number of algebraic equations. This is also called  \textit{by quadratures}. Let us show how this derivation goes, following \cite{Babelon:2003qtg}. \newline

      \begin{proof} 
       First let us recall that in a coordinate-free language, the existence of a non-degenerate Poisson bracket is equivalent to the existence of a non-degenerate two-form $\omega$ on $M$ which is closed, $d\omega = 0$, i.e.~a symplectic form.   For example, in the canonical coordinates one can show that $\{ F,G \} = \omega (X_F ,X_G)$   with $X_F$ the vector field associated to the function $F$, see \eqref{eq:MfVectors}, and
\begin{equation}
    \omega =  dp_i \wedge dq_i .
       \end{equation}       
       Note that in  canonical coordinates $\omega$ has constant components  and therefore $d\omega = 0$ holds trivially. In fact we have $\omega = d\alpha$ with
\begin{equation}
       \alpha = p_i dq_i ,
       \end{equation}       
        which is also known as the canonical one-form.

    \begin{exercise}
    Proof the above statements.
    \end{exercise}

\noindent Now suppose that \textit{on} $M_f$  we can solve for $p_i$ in terms of $F_i$ and $q_i$, i.e.~$p_i = p_i (F_i,q_i)$. This amounts to solving a ``finite number of algebraic equations" and may introduce some multi-valuedness in the problem (as we will illustrate with the harmonic oscillator). Then consider integrating the canonical one-form over a path $\gamma$ on $M_f$,
\begin{equation}
S \equiv \left.  \int_\gamma \alpha \right\vert_{F} \ .
\end{equation}
This is a well-defined function which does not depend on  the integration path $\gamma$, as we will now show. Consider two different paths $\gamma$ and $\gamma'$ going from the point $ ( p( F, q_0) , q_0  ) $ to $ ( p(F,q) ,q )$ in $M_f$.
\begin{figure}[H]
\centering
\includegraphics[scale=.36]{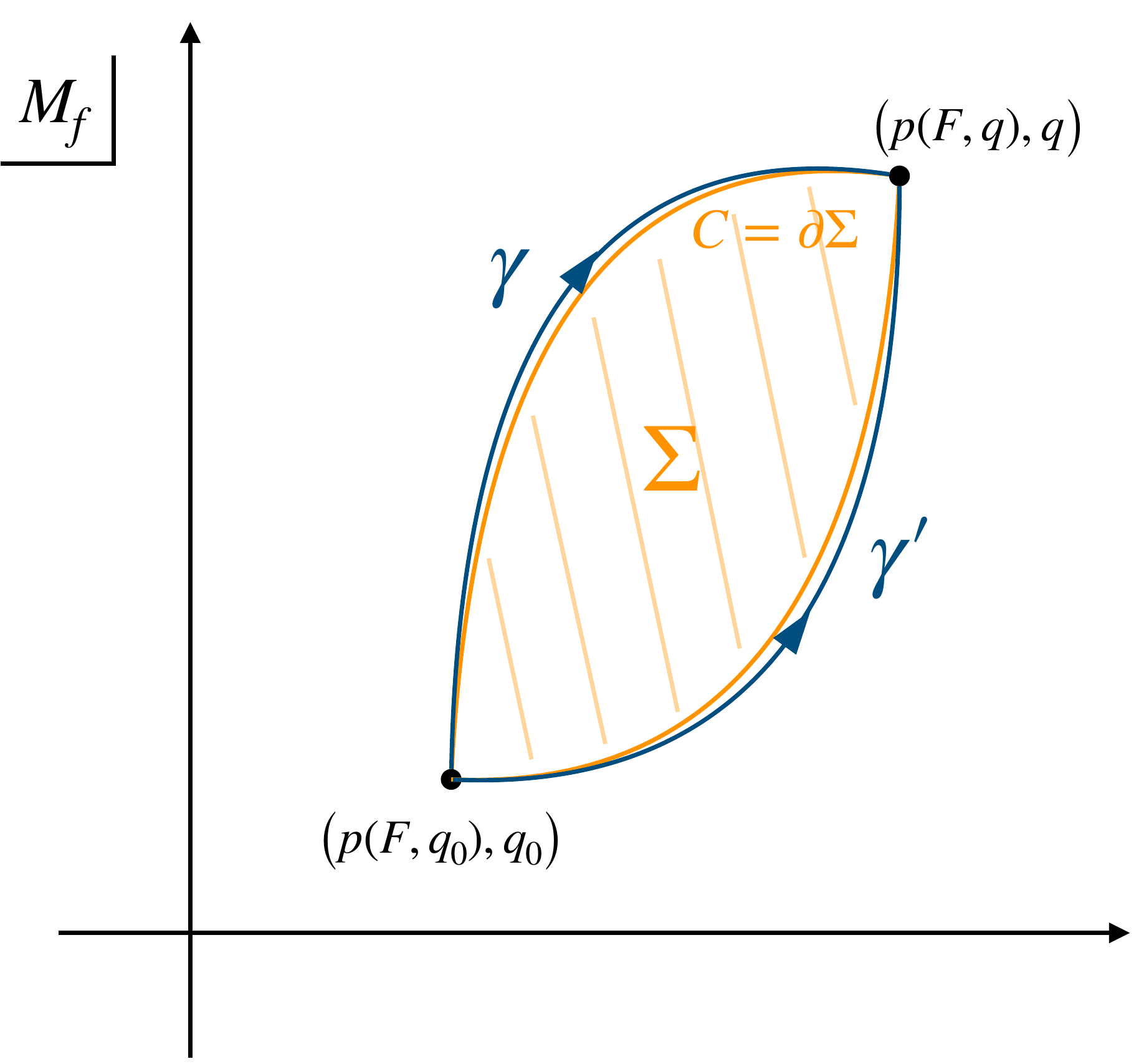}
\caption{Illustration to  support the proof that the function $S$ is independent of the chosen path $\gamma$.}
\end{figure}
\noindent Then, as can be seen from the above figure, we can define the surface $\Sigma$ with boundary $C = \partial \Sigma$ encapsulated by these two paths. Using Stokes' theorem we then find for the difference between the two functions $S(\gamma)$ and $S(\gamma')$  that
\begin{equation}
\delta S = S(\gamma) - S(\gamma')=\int_\gamma \alpha - \int_{\gamma'} \alpha = \oint_C \alpha = \oint_{\partial \Sigma} \alpha = \int_\Sigma d\alpha = \int_{\Sigma} \omega
\end{equation}
where we used $\omega = d\alpha$. In fact we have that $\omega\vert_{M_f} = 0$ because the conserved quantities $F_i$ are   in involution and independent which is equivalent to $dF_i (X_{F_j}) = \omega (X_{F_i} , X_{F_j}) =X_{F_j} (F_i)=\{F_i ,F_j \}=0 $. Hence, since $\Sigma \subset M_f$, we can conclude that $\delta S = 0$ and thus $S$ indeed exists and is (locally) well-defined.

Let us now take a particular path for $S$ so that we can view it as a well-defined function of $q$ as
\begin{equation}
     S (F_i,q_i) = \left. \int^q_{q_0} p_i (F_i , q'_i) \right\vert_F dq'_i  \ .
     \end{equation}     
Clearly $p_i = \left. \dfrac{\partial S}{\partial q_i} \right\vert_F$. If we further define $\psi_i \equiv \left.  \dfrac{\partial S}{\partial F_i} \right\vert_{q}$ then we can write
     \begin{equation}
     dS = \psi_i dF_i + p_i dq_i = \psi_i dF_i + \alpha ,
     \end{equation}
     and, therefore, we find
     \begin{equation}
     \begin{aligned}
     \omega &= d p_i \wedge q_i \\
      &= d\alpha = - d ( \psi_i d F_i ) = d F_i \wedge d \psi_i .
     \end{aligned}
     \end{equation}
In other words, we have shown that  there exists a canonical transformation $(p_i, q_i ) \rightarrow (F_i , \psi_i)$ to a coordinate system based on the conserved quantities. The generating function of this transformation is $S$. In this new set of canonical coordinates (if you wish ``adapted coordinates") the new momenta coincide with the constants of motion and the new coordinates are given by $\psi_i$.\footnote{Recall that the maximal number of conserved quantities is $n$.} Solving the EOMs then becomes trivial. In fact we have\footnote{Note that the following is consistent with $\dot{F}_i = - \frac{\partial H(F)}{\partial \psi_i} = 0$.}
\begin{equation}
\begin{aligned}
\dot{F}_i &= \{ H,F_i \} = 0,\\
\dot{\psi}_i &= \{ H , \psi_i \} = \dfrac{\partial H(F)}{\partial F_i} \equiv \Omega_i ,
\end{aligned}
\end{equation}
		where $\Omega_i $ by definition  is a constant in time and, therefore,  simple integration gives  explicit solutions linear in time, 
\begin{equation}
F_i (t) = F_i (0) , \qquad \psi_i (t) = \psi_i (0) +  \Omega_i t .
\end{equation}
In principle, we have therefore solved the system exactly (and in a very simple way).
However,  these new variables are not necessarily the most transparent ones. To reconstruct the dynamics of the original variables $(p_i,q_i)$ from these simple evolution equations we have to solve the ``inverse" problem, meaning that we need to invert  $\psi_i = \psi_i (F,q)$ to get $q_i = q_i (F, \psi )$ and consequently $p_i = p_i (F, q) = p_i (F, \psi)$ to obtain $(\dot{p}_i, \dot{q}_i)$ from $(\dot{F}_i , \dot{\psi}_i)$. Even though the system has been solved exactly, in practice this last (inverse) step can be in fact a difficult algebraic problem.
\end{proof}

Well known systems that are Liouville integrable include  the harmonic oscillator and  two-body central systems such as the Kepler problem, see e.g.~\cite{Torrielli:2016ufi}. Before discussing the (anisotropic) harmonic oscillator as an illustration, let us first discuss another set of useful variables to describe the motion on the level set manifold $M_f$.

\subsection{Action-angle variables} \label{s:AA}

Typically the $n$-dimensional  manifold $M_f$ has a non-trivial topology with non-trivial cycles along which the coordinates $\psi_i$ are multi-valued. When $M_f$ is compact and connected, Arnold-Liouville's theorem shows that it is diffeomorphic to an $n$-dimensional torus. This follows from the existence of a flat (integrable) connection, as implied by the commutation of the linearly independent basis vector fields $X_{F_i}$ of the tangent space of $M_f$ since
\begin{equation}
[X_{F_i}, X_{F_j}] = X_{\{F_i, F_j\}} = 0 .
\end{equation}
The so-called action-angle variables are coordinates that are adapted to the description of $M_f$ as an $n$-torus, and in this section we will construct them for completeness.

Let us consider the case that $M_f$ has $n$ non-trivial cycles $C_i$. Although $S$ is locally well-defined, i.e.~continuous deformations of the integration path do not affect $S$, globally the function $S$ will  be multi-valued along $C_i$. The \textit{action variables} $I_i$ are obtained by varying $S$ along these cycles,
\begin{equation}
I_i (F) =\frac{1}{2\pi} \oint_{C_i} d S \vert_{F} =\frac{1}{2\pi} \oint_{C_i} \alpha .
\end{equation}
They depend only on the conserved quantities $F_i$ and thus are constant on $M_f$. Let us assume that the action variables are also independent. We can therefore rewrite $S$ as a function $S = S(I, q)$. Then define the  variables $\theta_i$ conjugate to $I_i$
\begin{equation}
\theta_i \equiv \left. \frac{\partial S}{\partial I_i} \right\vert_q = \left. \frac{\partial S}{\partial F_j} \right\vert_{q}  \frac{\partial F_j}{\partial I} =  \psi_j \frac{\partial F_j}{\partial I_i} .
\end{equation}
They are  functions $\theta_i = \theta_i (\psi, I) = \theta_i (q, F)$.
On $M_f$ (on which $I_i$ is constant, so $dI_i \vert_{M_f} = 0$) they satisfy,
\begin{equation} \label{eq:proptheta}
\left. \oint_{C_j} d\theta_i \right\vert_I = \oint_{C_j} \left. \frac{\partial \theta_i}{\partial q_k} \right\vert_I dq_k = \oint_{C_j} \left. \frac{\partial^2 S}{ \partial I_i \partial q_k} \right\vert_F dq_k = \left. \oint_{C_j} \frac{\partial p_k}{\partial I_i}   \right\vert_I dq_k = \frac{\partial}{\partial I_i} \oint_{C_j} \alpha = 2\pi \delta_{ij} ,
\end{equation}
where we used interchangeably the fact that constant $F$ implies constant $I$, and  in the latter step we used the definition of $I_i$. The property \eqref{eq:proptheta} tells us that the $\theta_i$ are \textit{angle variables} which change $2\pi$ along the corresponding cycle $C_i$. The action-angle variables $(I_i, \theta_i)$ are canonical coordinates as well, 
\begin{equation}
\{ I_i , \theta_j \} = \frac{\partial I_i}{\partial F_k} \frac{\partial \theta_j}{\partial \psi_k} = \frac{\partial I_i}{\partial F_k} \frac{\partial F_k}{\partial I_j} = \delta_{ij} , \qquad \{ I_i , I_j\} = \{\theta_i , \theta_j \} = 0 \ ,
\end{equation}
and they satisfy a similar simple time-dependent system as the $(F_i , \psi_i)$ variables,
\begin{equation}
\dot{I}_i = - \frac{\partial H}{\partial \theta_i} = -\frac{\partial H}{\partial I_k} \frac{\partial I_k}{\partial \theta_j } -\frac{\partial H}{\partial \psi_k} \frac{\partial \psi_k}{\partial \theta_j }  =  0 , \qquad \dot{\theta}_i = \frac{\partial H}{\partial I_i} , \quad \text{with} \quad \ddot{\theta}_i = 0 .
\end{equation}
 As we mentioned, their merit is that they are adapted to the global description of $M_f$ as an $n$-dimensional torus. Under the evolution, the tori $M_f$ remain invariant and the motion is described by the angle coordinates $\theta_i$ dual to the cycles $C_i$. Note that nearby trajectories will  never exponentially diverge:  the evolution of an integrable system is so constrained and simple that it may never be chaotic.\footnote{Weak non-linear perturbations may spoil this nice confined behaviour, so that an integrable system may  become chaotic. The stability of the orbits on $M_f$ under these perturbations  is characterised by the Kolmogorov–Arnold–Moser (KAM) theorem.}

\subsection{A simple example: the anisotropic harmonic oscillator} \label{s:exampleHO}

A very simple example to illustrate the machinery behind Liouville's method is given by the  $d=1$-dimensional classical harmonic oscillator. We closely follow \cite{Torrielli:2016ufi} for this. The system  is described by the Hamiltonian (in units for mass $m=1$)
       \begin{equation}
       H = \dfrac{p^2}{2} + \dfrac{\omega^2 q^2}{2} ,
       \end{equation}
       and because $d=1$ it is the only conserved quantity we need for  classical integrability. 
       The foliation of the 2-dimensional phase space is given by ellipses on which $H=E$ is constant,
\begin{equation}
       M_f = M_E = \{ (p,q) \ | \ p^2 + 
       \omega^2 q^2 = 2 E = R^2 = \text{ constant} \} .
       \end{equation}       
       We now want to go to the coordinates adapted to these submanifolds.
       For convenience let us 
       %consider $R$ as our conserved quantity (instead of $H=E$) and already
        introduce the coordinates $(R,\phi)$ with $R>0$ as
       \begin{equation}
       q = \frac{R}{\omega} \cos\phi , \qquad  p = R \sin\phi ,
       \end{equation}
       which immediately relates $p$ to the conserved quantity $E$ (or $R$).
       For $q_0 = 0$ we have that
       \begin{equation}
       S (E,q) = - \dfrac{R^2}{\omega} \int^\phi_0 \sin^2 \phi' d\phi' = \dfrac{R^2}{2\omega} (\cos\phi\sin\phi - \phi) .
       \end{equation}
       To find $\psi = \left.\dfrac{\partial S}{\partial E}\right\vert_q$ we should be careful since we need to express $\phi = \phi (q,E)$. Since $\mathrm{arccos}$ only has $[0,\pi]$ as its domain, this  requires to distinguish  between $p>0$ and $p<0$.\footnote{Notice we would have the same problem concerning multi-valuedness when we would not have introduced $(R,\phi)$ and continued with $(p,q)$ instead. The we would have $p = p(E,q) = \pm \sqrt{2 E - \omega^2 q^2}$.} In particular, we have
       \begin{equation}
0<\phi<\pi : \quad      \phi = \arccos \frac{\omega q}{\sqrt{2E}} , \qquad \pi<\phi<2\pi :  \quad \phi= 2\pi - \arccos \frac{\omega q}{\sqrt{2E}} \ .
       \end{equation}
       Then, after obvious computations, we find
       \begin{equation}
       p\geq 0: \quad \psi = -\frac{1}{\omega}\arccos \frac{\omega q}{\sqrt{2E}} , \qquad p<0: \quad \psi =\frac{1}{\omega} \left(-2\pi +\arccos \frac{\omega q}{\sqrt{2E}} \right),
       \end{equation}
       and thus
       \begin{equation}
       \psi = -\frac{\phi}{\omega} .
       \end{equation}
       Since $\Omega$ is simply
       \begin{equation}
       \Omega = \frac{\partial H}{\partial E} = 1,
       \end{equation}
       the solution to the EOMs is
       \begin{equation}
       R(t) = R(0) = R, \qquad \psi (t) = \psi (0) +  t .
       \end{equation}
      
%       \begin{equation}
%       p(t) = c_1 \cos \omega t - c_2 \omega \sin \omega t , \qquad  q(t)  = c_2 \cos \omega t + \frac{c_1}{\omega} \sin \omega t
%       \end{equation}
       
%        We want to go to the coordinates adapted to these submanifolds. The solution for $p$ is
%       \begin{equation}
%       p = \pm \sqrt{ 2 H - \omega^2 q^2 } 
%       \end{equation}
%       and thus we have two branches $p>0$ and $p<0$ so that $S$ is multivalued.
%       \begin{equation}
%       S = 2 H q - \frac{\omega^2 q^3}{3} - S_0 , \qquad \psi = 2 (q-q_0)
%       \end{equation}
%    

    \begin{exercise}
  Perform the ``inverse problem" in order to obtain the evolution of the original variables. One should easily find  that
        \begin{equation} \label{eq:solaho}
         q(t)  =  \frac{\sqrt{2 E}}{\omega} \cos ( \omega t + \varphi) , \qquad p(t) = - \sqrt{2E} \sin ( \omega t + \varphi) , 
       \end{equation}
       for some initial phase $\varphi = \omega \psi (0)$. Show that \eqref{eq:solaho} indeed solves the EOMs of the anisotropic harmonic oscillator.
    \end{exercise}

For completeness, let us also calculate the action-angle variables, which is quite trivial here. We have for the cycle $C =\phi \in [0,2\pi]$ that the action variable is,
\begin{equation}
I = \frac{1}{2\pi} S_{\phi = 2\pi} = - \frac{E}{ \omega} ,
\end{equation}
while the angle variable is 
\begin{equation}
\theta = \left.\frac{\partial S}{\partial I}\right\vert_q = - \omega \left. \frac{\partial S}{\partial E}\right\vert_q = - \omega \psi = \phi .
\end{equation}
The action-angle variables correspond to the elliptic (polar) coordinates describing the ellipses of constant energy which foliate phase space.

The Kepler problem is another interesting example to discuss, since this system is known as a super-integrable system in which the number of independent conserved quantities exceeds the number of degrees of freedom (however, only $n=3$ quantities will Poisson commute amongst each other).  Unfortunately time does not permit to discuss the Kepler problem  but we refer the interested reader to \cite{Torrielli:2016ufi}.

\subsection{Lax pair formulation} \label{s:LaxPairFinite}

We have seen that the notion of Liouville integrability is  powerful: its generic definition and proof of solvability is model-independent, and it ensures
 a simple linear description of  a complicated classical system. However, it also suffers from a few drawbacks.
 Given a  particular classical dynamical system, in practice it is usually a very non-trivial task to determine whether it  is integrable or not. From scratch, there is no systematic way to find the $n$ independent conserved quantities in involution. Furthermore, generalisations such as solving  systems at the quantum level requires extra care, and  the concept of having as many conserved charges as degrees of freedom is no longer meaningful  when we want to describe classical \textit{field} theories, as in this case the number of degrees of freedom becomes infinite by the continuum limit.\footnote{How would you start to find and write down an \textit{infinite} number of conserved quantities?} A more modern approach, which in some sense addresses all these issues, is the (equivalent) Lax pair formulation of classical integrability. It gives a systematic way to construct conserved charges and describes a structure of classically integrable systems with a more apparent generalisation to field theories. In this section we will introduce the Lax pair formulation, and at the same time we are thus setting the stage for the next chapter.\\

\noindent {\bf Conserved charges} --- Suppose the equations of motion  can be recast in terms of two non-singular phase-space valued $n\times n$ square matrices $L$ and $M$  called the  \textit{Lax pair} as,
\begin{equation} \label{eq:LiouvilleLaxPair}
\partial_\tau L = [M,L] \ .
\end{equation}
An important remark is that a Lax pair is not unique. For instance one can perform a \textit{gauge} transformation $L \rightarrow g L g^{-1}$ and $M \rightarrow g M g^{-1} + \partial_\tau g g^{-1}$, with $g$ an invertible phase-space valued matrix, which leaves \eqref{eq:LiouvilleLaxPair} invariant.\footnote{In addition also the size of the Lax matrices may change, or we may change $L$ and $M$ with constant shifts $\alpha, \beta$ as $L \rightarrow L + \alpha\mathbf{1}$ and $M \rightarrow M + \beta \mathbf{1}$.} 

When a Lax pair is known, the construction of conserved quantities is very simple and follows  immediately from taking traces of powers of the matrix $L$ since,
\begin{equation}
\partial_\tau \Tr (L^m) = m \Tr ([M,L^m]) = 0 , \qquad \forall \, m \in \mathbb{N}_0 .
\end{equation}
Hence, we have a tower of conserved charges (which are not necessarily all independent),
\begin{equation}
F_m = \Tr L^m , \qquad \forall m \in \mathbb{N}_0.
\end{equation}
Equivalently, one can infer that the $n$ eigenvalues of $L$ are conserved. This can be easily seen using the gauge freedom. Suppose that $L$ can be diagonalised to a matrix $D$ as $L = U D U^{-1}$. The matrix $M$ must then transforms  to a matrix $C$ as $M = U C U^{-1} + \partial_\tau U U^{-1}$. The EOMs in terms of the Lax pair then take the form $\partial_\tau D = [C,D]$ and because the right hand side has no diagonal elements we have $\partial_\tau D_i = 0$ with $D_i$ the eigenvalues of $L$. \\

\noindent {\bf Involution of charges} --- Assuming that the conserved quantities are independent, Liouville integrability is not yet guaranteed, as the independent charges need to be in involution as well. In the Lax pair formulation, however, involutivity is ensured immediately
 when  the Poisson brackets of the theory  take a certain form. Before showing this, let us first introduce some notations. Note that the (phase-space valued) Lax matrices $L,M$ are elements of a Lie algebra $\mathfrak{g}$ (e.g.~simply $\mathfrak{g} = \mathfrak{gl}(n)$). We denote the generators of $\mathfrak{g}$ by $T_A$, $A = 1, \cdots , \dim \mathfrak{g}$ and we expand algebra elements as $X = X^A T_A \in \mathfrak{g}$ where the components $X^A$ are functions of the phase-space variables. Furthermore, we will consider maps from $\mathfrak{g}$ to elements of the tensor product $\mathfrak{g} \otimes \mathfrak{g}$ for which we introduce the notation,
\begin{equation}
X \mapsto X_1 \equiv X \otimes 1  , \qquad X \mapsto X_2 \equiv 1 \otimes X \ ,
\end{equation}
for $X\in \mathfrak{g}$.
 When there are $n$ tensor product factors we consider a similar notation, for example
 \begin{equation}
 X_3 = 1 \otimes 1 \otimes X \otimes 1 \otimes \ldots \otimes 1 \ .
 \end{equation} 
 Important for us will be the so-called $r$-matrix denoted by,
\begin{equation}
r_{12} = r^{AB} T_A \otimes T_B ,
\end{equation}
its permutation,
\begin{equation}
r_{21} = r^{BA} T_A \otimes T_B ,
\end{equation}
and the Casimir operator $C_{12}$,
\begin{equation} \label{eq:defc12}
C_{12} \equiv \eta^{AB} T_A \otimes T_B, 
\end{equation}
in which $\eta^{AB}$ is the inverse of a bilinear form on $\mathfrak{g}$ defined as $\eta_{AB} = \Tr (T_A T_B)$. Lastly, the Poisson bracket structure on the tensor product  takes the form,
\begin{equation}
\{ X_1 , Y_2 \} = \{ X^A , Y^B \} T_A \otimes T_B .
\end{equation}
Note that, in this notation, the Leibniz rule still takes the usual form,
\begin{equation}
\{ X_1 Y_1 , Z_2 \} =X_1 \{ Y_1 , Z_2 \} + \{ X_1 , Z_2\} Y_1 . \label{eq:tensorprop1} 
\end{equation}

    \begin{exercise}
 To get used to the notation, show the useful properties
\begin{align}
& [C_{12} , X_1] = - [C_{12} , X_2] , \\
& \Tr_2 (C_{12} X_2) = X_1 .
\end{align}
where $[\cdot, \cdot]$ corresponds to the usual Lie commutator and $\Tr_i$ corresponds to  taking the trace only on the $i$-th term of the tensor product.
    \end{exercise}

\noindent Using this notation one can show that 
the conserved eigenvalues of $L$ are in involution \textit{if and only if}
 the Poisson brackets between the $L$ matrix of the Lax pair takes the following simple form,
\begin{equation} \label{eq:LiouvilleInvolution}
\{ L_1 , L_2 \} = [r_{12}, L_1] - [r_{21} , L_2] ,
\end{equation}
for some $r$-matrix. \newline

\begin{proof}
\noindent {This proof follows \cite{Babelon:2003qtg}}. Consider again working in the ``diagonal gauge" $L = U D U^{-1}$. Using \eqref{eq:tensorprop1} one can show that
\begin{equation}
\{L_1 ,L_2 \} = U_1 U_2 \{ D_1 , D_2 \} U_1^{-1} U_2^{-1} + [t_{12}, L_1] - [t_{21}, L_2] ,
\end{equation}
where,
\begin{equation}
t_{12} = U_2 \{ U_1 , D_2 \} U_1^{-1} U_2^{-1} + \frac{1}{2} [\{U_1, U_2\}U_1^{-1} U_2^{-1}, L_2] . 
\end{equation}
$\bullet$ $\Rightarrow$ If the eigenvalues of $L$ are  in involution, then $\{D_1, D_2\} = 0$ and the Poisson brackets between the Lax matrices indeed take the form of eq.~\eqref{eq:LiouvilleInvolution} for $r_{12}=t_{12}$. \\
$\bullet$ $\Leftarrow$ Assuming eq.~\eqref{eq:LiouvilleInvolution} holds, then we have
\begin{equation}
\{ D_1 , D_2 \} = [s_{12} , D_1] - [s_{21}, D_2] ,
\end{equation}
with $s_{12} = U_1^{-1} U_2^{-1} (r-t) U_1 U_2$. Since the right hand side has zero on the diagonal, we can conclude that the eigenvalues of $L$ are in involution. 
\end{proof}

    \begin{exercise}
Show the statements of the proof above.
    \end{exercise}   

\vspace{.2cm}

\noindent For completeness, let us remark that under the gauge transformation $L \rightarrow g L g^{-1}$ and $M \rightarrow g M g^{-1} + \partial_\tau g g^{-1}$ the {form} of the EOMs and the Poisson brackets remain preserved, given that the $r$-matrix  transforms as
\begin{equation}
r_{12} \rightarrow g_1 g_2 \left( r_{12} + g_1^{-1} \{ g_1 , L_2 \} + \frac{1}{2} [g_1^{-1} g_2^{-1} \{g_1 , g_2\} , L_2] \right) g_1^{-1} g_2^{-1} .
\end{equation}
In addition, we always have the freedom to redefine the $r$-matrix as,
\begin{equation}
r_{12} \rightarrow r_{12} + [\sigma_{12}, L_2] ,
\end{equation}
with $\sigma$ a symmetric matrix, since this transformation leaves the Poisson bracket \eqref{eq:LiouvilleInvolution} invariant. \\

\indent Classifying in full generality  integrable  systems in the Lax pair formulation, which means providing the data of Lax pair and  $r$-matrix, is still an open and challenging problem. There are, however,  a few consistency relations which  determine large classes of $r$-matrices. For instance, when the $r$-matrix is constant and anti-symmetric $r_{12} = -r_{21}$ one can show  \cite{Babelon:2003qtg} that a sufficient condition for the Poisson bracket  in \eqref{eq:LiouvilleInvolution} to satisfy the Jacobi identity is that the $r$-matrix satisfies  the classical Yang-Baxter equation (CYBE),
\begin{equation} \label{eq:cYBE}
[r_{12} , r_{13}] + [r_{12}, r_{23}] + [r_{32}, r_{13}] = 0 .
\end{equation}
Alternatively, it is also sufficient that the $r$-matrix satisfies the modified Yang-Baxter equation (mCYBE),\footnote{These {constant classical $r$-matrices} (satisfying the CYBE or mCYBE) are in contrast to  dynamical (non-constant) $r$-matrices, which comprise a different class of $r$-matrices. }
\begin{equation} \label{eq:mYBE}
[r_{12} , r_{13}] + [r_{12}, r_{23}] + [r_{32}, r_{13}] = C_{12} ,
\end{equation}
where $C_{12}$ was defined in \eqref{eq:defc12}.
There is a lot of algebraic structure behind the solutions of these equations, and thus  $r$-matrices in general. They  therefore govern also  much information  on the quantum theory, and we will give a few more comments on them in section \ref{s:qint}.\\

Let us now briefly conclude on what we have seen in this section. When one has a Lax pair   satisfying \eqref{eq:LiouvilleLaxPair}, ensuring conserved quantities, and an $r$-matrix satisfying \eqref{eq:LiouvilleInvolution}, ensuring involutivity,  then the system is Liouville integrable.\footnote{As a curiosity one may wonder if for any Liouville integrable system, one can construct a Lax pair. This is in fact possible, and it is done by using the action-angle variables of section \ref{s:AA}, see e.g.~\cite{Babelon:2003qtg}.} These objects are very convenient to display an ``integrable structure" for the finite-dimensional classical systems. Nevertheless, also this formulation suffers from a few drawbacks, the biggest one being that it is in general hard to  guess or find the Lax pair, and it would be a major milestone when one is able to do so for a particular classical system. Indeed, a systematic approach to construct Lax pairs in full generality is lacking.\footnote{This is an active area of research. In the context of classical field theories, however, great development is being made, see e.g.~\cite{Vicedo:2017cge,Vicedo:2019dej,Lacroix:2020flf}. } Proving that a system is \textit{not} classically integrable is therefore not straightforward as well, and it is usually only done by showing that the dynamics exhibits chaotic behaviour. \\
%Furthermore, an important remark is that the lack of a Lax pair formulation for a certain system is \textit{not sufficient} to conclude that the system is \textit{not integrable}.

In the next chapter, we will generalise these ideas to classical field theories. With this in mind let us mention that in some interesting integrable systems one can actually find a one-parameter \textit{family} of Lax pairs satisfying \eqref{eq:LiouvilleLaxPair} since they depend  on a free \textit{spectral parameter} $z\in \mathbb{C}$. This is for example the case in the Kepler problem, see e.g.~\cite{Torrielli:2016ufi}. As we will see this situation generalises nicely to integrable field theories. 

\subsection{Once more: the anisotropic harmonic oscillator} \label{s:exampleHO2}

To illustrate the Lax pair formulation, let us go back to the anisotropic harmonic oscillator.  \\

    \begin{exercise}
Show that eq.~\eqref{eq:LiouvilleLaxPair} for the Lax pair,
\begin{equation}
L = \begin{pmatrix}
p & \omega q \\ \omega q & -p
\end{pmatrix} ,\qquad M =\frac{1}{2} \begin{pmatrix}
0 & \omega \\ -\omega & 0
\end{pmatrix} ,
\end{equation}
reproduces the EOMs of the anisotropic harmonic oscillator, and that $H = \frac{1}{4} \Tr L^2$.
    \end{exercise}

\noindent The $r$-matrix of this problem reads,
\begin{equation}
r_{12} = \frac{\omega}{4 H} \begin{pmatrix}
0 & 1 \\ -1 &0
\end{pmatrix} \otimes L .
\end{equation}
Note that this $r$-matrix $r^{AB}$ is dynamical and not anti-symmetric. \newline

    \begin{exercise}
Show that the Poisson brackets indeed satisfy eq.~\eqref{eq:LiouvilleInvolution} with the above $r$-matrix by using the fundamental Poisson brackets $\{p,q\} =1$.
    \end{exercise}   
    
    \newpage
    
\section{Classical integrability in two-dimensional field theories} \label{s:IFT}

In field theories, the number of degrees of freedom is infinite and a generalisation of the standard formulation of Liouville's integrability becomes rather inappropriate. As we will show,  the Lax pair formulation, however, turns out to be more adequate  to generalise.  We will consider  only  $(1+1)$-dimensional field theories for two main reasons. First,  higher-dimensional integrable field theories turn out to   be trivial (free) at the quantum level.\footnote{This statement can be understood from the fact that $S$-matrices of massive two-dimensional integrable field theories factorise into $2 \rightarrow 2$ elastic processes \cite{Zamolodchikov:1978xm,Parke:1980ki}. In higher dimensions  the corresponding result is that there are no non-trivial scatterings at all. This can be made precise by the Coleman-Mandula theorem \cite{Coleman:1967ad}.} Second, 
taking $d=2$ means that most of the concepts introduced here are also applicable to worldsheet sigma-models with applications to string theory. 

\subsection{Lax connection and the monodromy matrix in integrable field theories} \label{s:ClassicalIFT}

 Suppose  we have a  general $(1+1)$-dimensional local Lagrangian field theory in a spacetime  $\Sigma$,
%with a global symmetry group $G$. 
which we will either define  on the infinite plane $\Sigma = \mathbb{R}^{1,1}$ or on the cylinder $\Sigma = \mathbb{R} \times S^1$  (the latter  with an eye on string theory applications). We denote the  time-direction with $\tau$ and the spatial-direction with $\sigma$. Suppose moreover one can recast the equations of motion  in a way similar to \eqref{eq:LiouvilleLaxPair} by means of two matrices ${\cal L}_\sigma(\tau, \sigma ; z)$ and ${\cal L}_\tau(\tau, \sigma ; z)$  depending now on a free \textit{spectral parameter} $z\in \mathbb{C}$  as  \cite{Zakharov:1973pp},
\begin{equation} \label{eq:FTLaxPair} 
\partial_\tau {\cal L}_\sigma (z) - \partial_\sigma {\cal L}_\tau (z) + [{\cal L}_\tau (z) , {\cal L }_\sigma (z) ] = 0 , \qquad \forall z\in \mathbb{C} ,
\end{equation}
where for brevity we have dropped the explicit $(\tau , \sigma)$ dependency. This condition is  called the \textit{zero-curvature} or \textit{flatness} condition and the matrices  ${\cal L}_\sigma(z)$ and ${\cal L}_\tau(z)$ are Lax pairs, as before. With the existence of the spectral parameter  they now take values  in the loop algebra $\mathfrak{g} \otimes \mathbb{C}$. 
%associated to the Lie symmetry group $G$. 
The condition \eqref{eq:FTLaxPair} can also be recast in a coordinate-independent way by introducing the \textit{Lax connection} one-form on $\Sigma$
\begin{equation}
{\cal L} (z) = {\cal L}_\tau(z) \, d\tau + {\cal L}_\sigma(z) \, d \sigma ,
\end{equation}
so that \eqref{eq:FTLaxPair} becomes,
\begin{equation}\label{eq:LaxZeroCurv}
d \mathcal{L}(z) + \mathcal{L}(z) \wedge \mathcal{L}(z) = 0 \, \qquad \forall z\in \mathbb{C}  .
\end{equation}
 Although there is no universal definition, we will define a \textit{weakly} classical integrable field theory as one where the EOMs can be represented through
 a flat Lax connection ${\cal L}(z)$ for any value of the spectral parameter $ z\in \mathbb{C}$. We will also call this the ``zero-curvature formulation".
The freedom in the parameter $z$ is crucial and in fact allows us to construct  \textit{infinite} sets of conserved charges or symmetries.  In this case we will briefly argue that there are certain methods, analogously to Liouville's quadratures, which render these field theories solvable.
As we will see later, to ensure that a certain infinite set of the conserved charges are further in involution, we will again have to require that 
 the Poisson brackets of the Lax matrices  take a certain form.   We then say that the field theory is also \textit{strongly} classically integrable.  \\
%
%Let us however remark that this definition of classical integrability in the field theory should not be confused with Liouville integrability. \newline

In the remainder of this section, we will  show how one can construct  infinite sets of conserved charges in the zero-curvature formulation.  First, let us note that the zero-curvature condition \eqref{eq:FTLaxPair} corresponds  to a compatibility condition of the following auxiliary linear system,
\begin{equation}
\left( \partial_\sigma + {\cal L}_\sigma (z) \right) \Psi (\tau , \sigma; z) = 0 , \qquad \left( \partial_\tau + {\cal L}_\tau (z) \right) \Psi (\tau , \sigma ; z)  = 0 ,
\end{equation}
where $\Psi (\tau , \sigma; z) $ is sometimes called the ``the wave-function" because of the resemblance with a time-dependent Schr\"odinger problem. The wave-function is fixed by the initial condition $\Psi (0,0 ; z) = 1$.
%\footnote{Let us point out that the auxiliarly linear problem is very useful when describing solitonic solutions.} 
One obtains its solution  by   parallel transporting   from the origin to a point $(\tau, \sigma)$ along an arbitrary path $\gamma$ with the connection ${\cal L}(z)$, which may be done unambiguously since ${\cal L}(z)$ is a flat. In other words,
\begin{equation}
\Psi ( \tau , \sigma ; z) =   \overleftarrow{P \exp} \left( - \int_\gamma {\cal L} (z) \right) ,
\end{equation}
 is well-defined (it does not depend on the path $\gamma$) and here $ \overleftarrow{P \exp}$ is  the path ordered exponential defined on a fixed time slice  as,
%\begin{equation}
\begin{align}
 \overleftarrow{P \exp} \left( \int_0^\sigma d\sigma' \, A(\sigma')  \right)  &= \sum_{n = 0}^\infty \frac{1}{n!} \int^\sigma_0 \cdots \int^\sigma_0   \overleftarrow{P } \{ A(\sigma_1') \cdots A(\sigma_n') \} \, d\sigma_1' \cdots d\sigma_n' \\
 &=  \sum_{n = 0}^\infty \int^\sigma_0 d\sigma_n' \int^{\sigma_n'}_0 d\sigma_{n-1}'  \cdots \int^{\sigma_2'}_0 d\sigma_1'  \ A(\sigma_n') \cdots A(\sigma_1')   \, . \nonumber
\end{align}
%\end{equation}
 An infinite set of conserved charges can now be obtained by considering a path at a fixed time slice to define the \textit{transport matrix} $T(b,a;z) \equiv \Psi (\tau , b ; z) \Psi (\tau , a ; z)^{-1}$,
\begin{equation} \label{eq:monodromy}
T(b,a; z ) = \overleftarrow{P \exp} \left(- \int^b_a \mathrm{d}\sigma\; \mathcal{L}_\sigma ( z) \right)\, .
\end{equation}

    \begin{exercise}
Write out the first three terms of the the expansion of $T(b,a;z)$. Note that in general this is a highly non-local expression.
    \end{exercise}

\vspace{.1cm}

\noindent The transport matrix satisfies the following  properties,
\begin{align}
\delta T(b,a ;z) &= - \int^b_a \mathrm{d}\sigma\, T(b,\sigma;z) \delta \mathcal{L}_\sigma (\tau, \sigma ; z) T(\sigma, a;z)\, ,\\
\partial_\sigma T(\sigma,a ;z) &= - \mathcal{L}_\sigma (\tau,\sigma ;z) T(\sigma, a;z)\, , \label{eq:Transport2}\\
\partial_\sigma T(b,\sigma ;z) &= T(b, \sigma ;z)  \mathcal{L}_\sigma (\tau,\sigma ;z)\, , \label{eq:Transport3}\\
T(a,a ;z ) &=1\,  \label{eq:Transport4},
\end{align}
Using the flatness of the Lax $\mathcal{L}(z)$ together with the above properties, one can  show that,
\begin{equation}\label{eq:MonToTime}
\begin{aligned}
\partial_\tau T(b,a ;z) 
%&=- \int^b_a \mathrm{d}\sigma\, T(b,\sigma;z) \partial_\tau \mathcal{L}_\sigma (\tau, \sigma ; z) T(\sigma, a;z)\, ,\\
%&=- \int^b_a \mathrm{d}\sigma\, T(b,\sigma;z)\left( \partial_\sigma \mathcal{L}_\tau (\tau, \sigma ; z)  - [\mathcal{L}_\tau (\tau, \sigma ; z) , \mathcal{L}_\sigma (\tau, \sigma ; z)]  \right)T(\sigma, a;z)\, ,\\
%&= - \int^b_a d\sigma \partial_\sigma \left[ \overleftarrow{P \exp} \left( \int^b_s ds {\cal L}_\sigma (s, t;z) \right) \mathcal{L}_\tau (\tau, \sigma ; z)  \overleftarrow{P \exp} \left( \int^s_a ds  {\cal L}_\sigma (s, t;z)  \right)  \right] , \\
&= T(b,a ;z) \mathcal{L}_\tau (\tau, a;z)  - \mathcal{L}_\tau (\tau, b ;z) T(b,a;z) \,  .
\end{aligned}
\end{equation}

    \begin{exercise}
Show  \eqref{eq:MonToTime}.
    \end{exercise}   

\vspace{.1cm}    

\noindent Let us now distinguish two cases.
\begin{itemize}
\item When $\Sigma = \mathbb{R}^{1,1}$ we typically impose asymptotic fall-off boundary conditions on our fields and the Lax pairs, i.e.~${\cal L}(\sigma \rightarrow \pm \infty ) \rightarrow 0$. In this case we  have
 that (any power of) the \textit{monodromy matrix} $T( +\infty , -\infty ; z)$  is trivially conserved,
\begin{equation}
\partial_\tau T( +\infty , -\infty ; z)^n = 0 ,
\end{equation}
for any $n \in \mathbb{N}$ and for any $z\in \mathbb{C}$. 
\item  When $\Sigma = \mathbb{R} \times S^1$ we impose periodic boundary conditions $\sigma \simeq \sigma + 2\pi$ in the spatial coordinate, and thus we will have ${\cal L} (\sigma + 2\pi) = {\cal L} (\sigma)$. Then
\begin{equation} \label{eq:MonTimePeriodic}
\partial_\tau T(2\pi,0 ;z) = [T(2\pi,0 ;z), \mathcal{L}_\tau (\tau, 0;z)],
\end{equation}
and thus
 it is the trace of (any power of) the \textit{monodromy matrix} $T(2\pi , 0; z)$ that is conserved:
\begin{equation}
\partial_\tau \Tr T(2\pi , 0 ;z)^n = 0 ,
\end{equation}
for any $n \in \mathbb{N}$ and for any $z\in \mathbb{C}$. Note that \eqref{eq:MonTimePeriodic} has the same form as the EOMs in terms of  Lax pairs  for finite-dimensional systems given in eq.~\eqref{eq:LiouvilleLaxPair} (up to a minus sign convention on ${\cal L}(z)$). To make a comparison one can think of the monodromy matrix $T$ as the Lax matrix $L$ and ${\cal L}_\tau$ as the Lax matrix $M$.
\end{itemize}
In the rest of these notes we will write $T(z)$ to denote the monodromy matrix $T( +\infty , -\infty ; z)$ or $\Tr T(2\pi , 0 ;z)$ depending on the boundary conditions taken.
 Now Taylor expanding  $T(z)$ around  suitable values of $z$ thus finally leads us to an infinite set of conserved charges. For example, when $T(z)$ is analytical around $z=0$ we can write
 \begin{equation}
 T(z) = \sum_{i=0}^\infty Q_i z^i , \quad \text{with} \quad \partial_\tau Q_i = 0 , \ \ \forall i \in \mathbb{N} \ .
 \end{equation}
  In this way, one can in fact obtain multiple infinite sets of conserved charges with different properties. Typically, there are two sorts of charges at play, local (higher-spin) charges and non-local charges; see e.g.\ \cite{Babelon:2003qtg,Loebbert:2016cdm}. We will discuss both of them explicitly  in section \ref{s:PCMint} for the Principal Chiral Model. In the rest of this section we briefly discuss an important set of local charges obtained by expanding $T(z)$ around poles of the Lax connection. Before doing so, let us make three final important remarks:
 \begin{itemize}
 \item  The charges obtained from expanding $T(z)$ are not necessarily all in involution: they can generate additional \textit{hidden} classical symmetry algebras under the Poisson brackets, as we will mention also later.
 \item  The Lax connection ${\cal L}(z)$ is again not unique; it is defined up to a local gauge transformation  given by,
\begin{equation}\label{eq:LaxGauge}
\mathcal{L}(z) \;\rightarrow \; \mathcal{L}^g(z) = g \mathcal{L}(z) g^{-1} - d g g^{-1}\, ,
\end{equation}
which leaves the zero curvature condition \eqref{eq:FTLaxPair} invariant. The (matrix) elements $g$ are completely arbitrary and can for instance also depend on the spectral parameter.  
Under the gauge transformation \eqref{eq:LaxGauge} the transport matrix transforms as,
\begin{equation} \label{eq:MonoGauge}
T(b,a;z) \; \rightarrow \; T^g(b,a ;z) = g(\tau,  b) T(b,a ;z ) g^{-1}(\tau, a) \, .
\end{equation}
\item Either from observing that the gauge transformations can be used to diagonalise the monodromy matrix, or from the possibility of taking any power $n\in \mathbb{N}$, we can conclude that the (collection of) eigenvalues $\lambda (z)$ of  $T(2\pi, 0 ;z)$ are also constant in time, and one may instead expand those in $z$ to obtain an infinite tower of conserved charges.
 \end{itemize}
 
 \vspace{.2cm}
    \begin{exercise}
Show \eqref{eq:MonoGauge}.
    \end{exercise}   
    
\vspace{.3cm}

\noindent \textbf{Local charges from {abelianization}} --- We end the discussion on constructing conserved charges in the zero-curvature formulation by showing how in general we may find an interesting set of \textit{local} conserved charges by expanding around poles of the Lax connection. This generic procedure is also known as \textit{abelianization}.

Let us assume that the Lax connection has a number of poles at constant values  $z_k$ of order $n_k$. In the vicinity of each pole $z_k$ one can perform a \textit{diagonal gauge} transformation to a  Lax connection ${\cal L}^{(z_k)} (z)$ which is diagonal by,
\begin{equation} \label{eq:DiagonalGauge}
{\cal L}^{(z_k)} (z) = g^{(z_k)}( z ) {\cal L}(z) g^{(z_k)}( z ){}^{-1} - d g^{(z_k)}( z ) g^{(z_k)}( z ){}^{-1} ,
\end{equation}
using matrices that are regular at the poles,
\begin{equation} \label{eq:DiagGaugeElement}
g^{(z_k)}( z ) = \sum_{n = 0}^\infty g_n^{(z_k)} (z- z_k)^n .
\end{equation}
The punchline of the proof given for example in \cite{Babelon:2003qtg} is that the matrices $g_n^{(z_k)}$ have enough freedom to diagonalise ${\cal L}(z)$ order by order in $z$ in a recursive way.
%{\color{red} complete this?}
 We can then expand ${\cal L}^{(z_k)} (z)$ around $z = z_k$ as,
\begin{equation} \label{eq:DiagLax}
{\cal L}^{(z_k)} (z) = \sum_{n= - n_k}^{\infty} {\cal L}_{n}^{(z_k)} (z - z_k)^n ,
\end{equation}
and the flatness condition \eqref{eq:LaxZeroCurv} of the diagonalised Lax connection  reduces to a local conservation law of the diagonalised coefficients $\star {\cal L}_{n}^{(z_k)} $,
\begin{equation} \label{eq:DiagZeroCurv}
d {\cal L}_{n}^{(z_k)}  = 0, \qquad n \geq -n_k .
\end{equation}
Without actually referring to the monodromy matrix this implies immediately the conservation of the following local set of  charges,
\begin{equation} \label{eq:localdiagonal}
Q^{(z_k)}_{n} = \int_\Sigma d\sigma \,   {\cal L }^{(z_k)}_{n}{}_\sigma , \qquad n \geq -n_k ,
\end{equation}
  under both periodic and asymptotic boundary conditions.
   They  do however appear in the \textit{diagonal} gauge transformed monodromy matrix (where the path ordering can be dropped) as, 
\begin{equation} \label{eq:DiagMono}
T^{g^{(z_k)}} (z) = \exp \left( - \sum_{n = -n_k}^\infty Q^{(z_k)}_{n} (z - z_k )^n \right) ,
\end{equation}
where we have performed \eqref{eq:MonoGauge} with the element $g^{(z_k)}(z)$ given in \eqref{eq:DiagGaugeElement}. \newline

    \begin{exercise}
Show that the local charges~\eqref{eq:localdiagonal} are indeed conserved for periodic and asymptotic boundary conditions. Compare this situation to a conserved  charge that is associated to a local conservation law of the form $d\star J = 0 $ (instead of $dJ = 0$) for $J$ a one-form on $\Sigma$ and the convention $\star d\tau = d\sigma$ and $\star d\sigma = d\tau$ for the Hodge star operation.
    \end{exercise}

%{\color{red}abelianisation gives local expression for wavefunction. nice for classical inverse scattering or finite gap eqs }

\subsection{Classical integrable methods} \label{s:ClassMethods}

While the zero-curvature formulation only ensures us that we have infinite towers of conserved charges, and does not ensure anything about their involution, one may already say that we possess the ``classical integrable structure" of the theory, meaning
\begin{equation}
\{ {\cal L}(z) , \ \Psi (z) , \ T(z) \} \ .
\end{equation}
The reason is that these objects form the backbone of several well-known classical integrable methods which do not need properties of the Poisson bracket structure. At the quantum level, however, the knowledge of the Poisson bracket algebra does play a very important role, as we will briefly discuss in section \ref{s:qint}. Here we will only give some general comments and sketch the punch lines of two particular classical methods known as  the Classical Inverse Scattering Method (CISM) and the Classical Spectral Curve (CSC) or finite-gap integration technique. \\

\noindent{\bf Classical Inverse Scattering} --- To solve the equations of motion (which are typically non-linear coupled partial differential equations) of an integrable field theory an established method worked out in the  60-70's is the Classical Inverse Scattering Method \cite{Gardner:1967wc,Zakharov:1979zz}. In  philosophy, the CISM  is  very analogous to performing a Fourier transformation  in order to solve a highly non-linear system. The idea is to interpret the equation $\partial_\sigma \Psi =- {\cal L}_\sigma \Psi$ of the auxiliary linear problem as a \textit{direct scattering} problem, where the wave-function $\Psi$ scatters off a potential ${\cal L}_\sigma$. One may then derive the scattering data such as the reflection and transmission coefficients which can be thought of as the \textit{auxiliary variables} (and in some sense are also seen as generalisations of the action-angle variables of Liouville integrability) and which will depend on the spectral parameter $z$. Because the potential ${\cal L}_\sigma$ is derived from a flat connection, these auxiliary variables typically will have a simple time evolution. This step in particular is only true in an integrable field theory. The \textit{inverse problem} is then a scattering transformation problem where one reconstructs the solution in terms of the original canonical field variables, depending on $\sigma$ and $\tau$, from the knowledge of the scattering data, depending on $z$ and $\tau$. Performing the inverse problem is usually hard, and it does not need any knowledge on the classical integrable structure. Let us point out that this is precisely what we eluded to in the introduction: the merit of classical integrability is the \textit{possibility} of {using}  methods to solve the system. In this case, we can in particular derive the evolution of the scattering data easily. Having such methods does not mean that we can, in fact, actually solve the problem. In this case,  it does not mean that it is analytically possible to solve or interpret the inverse problem. The ability to accomplish the latter is highly model-dependent.  For an illustration of the CISM to solve the Korteweg-De Vries equation or the Sine-Gordon equation, see for instance \cite{Babelon:2003qtg,Torrielli:2016ufi}. \\

\noindent{\bf Classical Spectral Curve} --- A more modern approach to solving classical integrable field theories is the Classical Spectral Curve or the finite-gap integration method. Rather than solving the equations of motion, the CSC provides directly expressions for the spectrum of charges of the theory, such as the energy, and relations between them, which at the end of the day is what one is in fact after. The large idea is again to start from the auxiliary linear problem
\begin{equation}
d\Psi = - {\cal L} \Psi \ ,
\end{equation}
and interpret it as a Dirac equation with a periodic potential ${\cal L}$.\footnote{We consider here $\Sigma = \mathbb{R} \times S^1$.} The wave-function $\Psi$, however, is quasi-periodic as we know from the definition of the transport matrix, namely we have 
\begin{equation}
\Psi (\sigma = 2\pi ) = T(2\pi , 0 ; z) \Psi (\sigma = 0) \ . 
\end{equation}
The spectrum for the wave-function will therefore have a band structure which is determined by the so-called \textit{quasimomenta} $p_i(z)$, which are related to the eigenvalues of $T(2\pi , 0 ; z)$. The core of the CSC method is that it  provides a complete analytical analysis of $p_i(z)$ on the full complex plane, and therefore of the full spectrum. 

Let us take a step back. As remarked in the previous section, the collection of eigenvalues $\lambda(z)$ of $T(2\pi,0;z)$ are constant in time and can be expanded in $z$ to obtain a tower of conserved charges. They are defined by the algebraic Classical Spectral Curve $\Gamma$, which simply corresponds to the characteristic equation
\begin{equation}
\Gamma = \det \left[ \lambda(z) \mathbf{1} - T(2\pi, 0 ;z) \right] = 0 \ .
\end{equation}
For a representation of the algebra in which $T(2\pi, 0 ;z)$ is an $M\times M$ matrix,   the characteristic equation will in general be a polynomial equation of degree $M$. Solving it will therefore  introduce an $M$-sheeted Riemann surface which may be connected through several branch cuts of which the branch points correspond to points in the complex $z$-plane where two or more of the eigenvalues degenerate.
 The location, size and number of these branch cuts depends on the particular solution to the equations of motion. From a different perspective, the data of the cuts can thus be taken as a way to classify families of solutions. To continue, one usually considers only  the class  marked by a finite number of branch cuts of finite size, and which are of square-root type (i.e.~they only connect two sheets). This is the reason that the CSC method also goes under the name of ``finite-gap integration technique".
 
 The study of the Riemann surface encoded by the eigenvalues $\lambda(z)$ is complicated by the fact that the Lax connection typically has a number of poles which introduce essential singularities for the monodromy matrix $T(2\pi, 0 ;z)$, and thus make  the curve highly singular. To simplify this problem, one introduces the so-called quasimomenta $p(z)$ as
 \begin{equation} \label{eq:DefQM}
 \lambda_i(z) = e^{i p_i(z)} \ ,
 \end{equation}
where $i$ runs over all the eigenvalues, $i = 1, \ldots , M$. The quasimomenta will then have the same pole structure as the Lax connection, as can be seen from \eqref{eq:DiagLax} and \eqref{eq:DiagMono}. Along the cuts, whose union between sheet $i$ and sheet $j$ is denoted by ${\cal C}_{(i,j)}$, the quasimomenta now  degenerate up to integer shifts in $2\pi $, that is
\begin{equation}
p_i (z + i 0 ) = p_j (z - i 0) + 2\pi n_{i,j} , \qquad n_{(i,j)} \in \mathbb{Z} \ ,
\end{equation}
for $z \in {\cal C}_{(i,j)}$. These equations are    Riemann-Hilbert problems which are also known as the finite-gap  or classical Bethe equations. When fixing $n_{(i,j)}$, as well as  knowing the asymptotics of the  quasimomenta, one usually has enough information to reconstruct $p_i(z)$ on the whole complex plane,\footnote{This is in particular possible when the quasimomenta have the following asymptotics \cite{Dorey:2006zj,Vicedo:2008ryn}
\begin{equation}
\begin{alignedat}{2}
p_i(z) & \approx p_i(0) + z p_i' (0) + {\cal O}(z^2) , \qquad && \text{around } \  z=0 \ , \\
p_i(z) & \approx \frac{p_i({\infty})}{z} + {\cal O}(z^{-2}) , \qquad && \text{around }  \  z=\infty \ ,
\end{alignedat}
\end{equation}
as is for instance the case for the Principal Chiral Model (PCM) of section \ref{s:PCM}. \label{f:QMAsympt}
} and therefore all the conserved quantities (the semiclassical spectrum) contained in $\lambda(z)$, see e.g.~\cite{Dorey:2006zj,Vicedo:2008ryn}.

Let us finally mention that, because the quasimomenta have poles and thus are not analytic, one often defines the so-called resolvent $G(z)$ in which the poles are simply removed 
\cite{Kazakov:2004qf,Beisert:2004ag,Beisert:2005bm}.
The resolvent is then analytic everywhere in the complex plane with cuts and can therefore be represented in terms of a density function. The finite-gap equations or classical Bethe equations are usually written as integral equations for these densities, and they can be interepreted as the continuum limit of their quantum counterparts.
For more details on the CSC and finite-gap integration method we refer to \cite{Dorey:2006zj,Vicedo:2008ryn} and the overview \cite{Zarembo:2010yz}.\footnote{See also the recent paper \cite{Borsato:2021fuy} in which a review on the CSC is included and in which the reconstruction of one-cut solutions for certain integrable deformations, namely  the homogeneous Yang-Baxter deformations, is performed.} Pioneering papers that used the CSC for AdS/CFT integrability, in particular to match the semiclassical spectrum of (sectors of) the AdS$_5\times S^5$ background with $N=4$ Super-Yang-Mills, are \cite{Kazakov:2004qf,Kazakov:2004nh,Beisert:2004ag,Schafer-Nameki:2004utl,Beisert:2005bm}  and they are also  good resources  to show how in fact one may reconstruct the quasimomenta in practice.

%Let us remark it is also possible to define  a \textit{twisted} transport matrix,
%\begin{equation} \label{eq:GenMon}
%T^\W(b,a; z ) = \overleftarrow{P \exp} \left(- \int^b_a \mathrm{d}\sigma\; \W\left[ \mathcal{L}_\sigma ( z)\right] \right)\, ,
%\end{equation}
%where $\W$ is a constant Lie algebra automorphism.
% %Given these properties of $\Omega$ one can show 
%Under time derivation it behaves as,
%\begin{equation}\label{eq:GenMonToTime}
%\partial_\tau T^\W (b,a ;z) =  T^\W(b,a ;z)\W\left[ \mathcal{L}_\tau (\tau, a;z) \right]  - \W\left[ \mathcal{L}_\tau (\tau, b ;z) \right] T^\W(b,a;z) \, , \\
%\end{equation}
%such that $\partial_\tau T(+\infty , -\infty ; z)^n = 0$ or $\partial_\tau \Tr T^{\W}(2\pi,0;z)^{n} = 0 $ for all $n\in \mathbb{Z}$ and $z\in \mathbb{C}$. Under gauge transformations we have,
%\begin{equation}\label{eq:GenMonGaugeTransf}
%T^\W(b,a;z) \; \rightarrow \; \omega\left( g( b) \right) T^\W (b,a ; z) \omega\left(g^{-1}(a)  \right) \, ,
%\end{equation}
%  where the map $\omega: G \rightarrow G$ is defined as $\omega \left(e^{tX}\right) = e^{t \W[X]} $ for $t$ small and $X\in \mathfrak{g}$, assuming the Lie group $G$ is connected to the identity (in this case $\omega$ is a constant Lie \textit{group} automorphism also). 
%  %This possibility has been vital for the work \cite{Driezen:2018glg} described in chapter \ref{Chapter10}.

\subsection{Charges in involution} \label{s:FTInvolution}

To have a \textit{strongly} classical integrable field theory we do not only require a zero-curvature formulation, but we also require 
that an infinite tower of the conserved charges is in involution. In particular, this can be ensured when the equal-time Poisson brackets between the spatial Lax components ${\cal L}_\sigma (z) $ takes a certain form similar to \eqref{eq:LiouvilleInvolution}. Contrary to the finite-dimensional case, which involved an ``if-and-only-if" statement,  here there are in principle several possibilities for the form of the  Poisson brackets $\{ {\cal L}_{\sigma, 1} (\sigma ; z) , {\cal L}_{\sigma , 2} (\sigma ' , z' \}$.\footnote{Note that here the Poisson bracket is taken between Lax matrices at different points $\sigma$ and $\sigma'$ and at different values of the spectral parameters $z$ and $z'$. Let us recall also that in a field theory, the Poisson brackets are defined as
\begin{equation}
\{ A (\sigma ) , B(\sigma') \} = \int d\sigma'' \left( \frac{\delta A (\sigma)}{\delta \pi_\mu (\sigma '')} \frac{\delta B(\sigma')}{\delta x^\mu (\sigma'') } - \frac{\delta B (\sigma')}{\delta \pi_\mu (\sigma '')} \frac{\delta A(\sigma)}{\delta x^\mu (\sigma'') }    \right) ,
\end{equation}
for the canonical fields $x^\mu (\sigma, \tau)$ and their conjugate momenta $\pi_\mu(\sigma , \tau) $ with $\mu = 1, \ldots , $ the number of fields.
} There are however only two main cases studied in the literature, as we will discuss. Note, nevertheless, that the existence of other structures are not necessarily excluded.\\

\noindent $\bullet$ In the  same notation as used in section \ref{s:LaxPairFinite}, 
when there exists an $r$-matrix such that\footnote{One should be really careful when comparing to other references where the convention on the Lax flatness condition \eqref{eq:FTLaxPair}  might contain an additional minus sign in front of the commutator. In that case one should send in all formulas that follow ${\cal L} \rightarrow -{\cal L}$.}
\begin{equation} \label{eq:FTInvolution1}
\{ {\cal L}_{\sigma, 1} (\sigma ; z) , {\cal L}_{\sigma , 2} (\sigma ' , z' \} =   \left[{\cal L}_{\sigma, 1} (\sigma, z) +{\cal L}_{\sigma, 2} ( \sigma, z') , r_{12}(z,z')  \right] \delta(\sigma - \sigma' )  ,
\end{equation}
then the involution of charges can be shown.
These Poisson brackets are called \textit{ultralocal}, which means that they only involve delta-functions, and not the derivatives of delta-functions.  To ensure they satisfy the Jacobi identity  it is, similarly as before, sufficient that  the $r$-matrix is skew $r_{12} (z,z') = - r_{21} ( z' , z)$, constant (in the fields) and satisfies the spectral parameter dependent CYBE,
\begin{equation} \label{eq:FTcYBE}
\left[ r_{12}(z_1, z_2) ,  r_{13}(z_1, z_3) \right] + \left[ r_{12}(z_1, z_2) ,  r_{23}(z_2, z_3) \right] + \left[ r_{32}(z_3, z_2) ,  r_{13}(z_1, z_3) \right] = 0 .
\end{equation}
From now on we will always assume that the $r$-matrix is constant in the fields. 
Using \eqref{eq:FTInvolution1} the Poisson brackets between transport matrices  take the form,
\begin{equation}
\{ T_1 (b,a ;z) , T_{2}(b,a ; z') \} = [r_{12}( z,z' ) ,  T_1 (b,a ;z)  T_{2}(b,a ; z')  ]  .
\end{equation}
This expression is known as the \textit{Sklyanin Exchange relations} \cite{Sklyanin:1980ij}. For a proof see \cite{Faddeev:1987ph} or the following exercise. \newline

    \begin{exercise}
Proof the Sklyanin Exchange relations from \eqref{eq:FTInvolution1} and using the definition of the path-ordered exponential as well as the properties \eqref{eq:Transport2}, \eqref{eq:Transport3} and \eqref{eq:Transport4}.
    \end{exercise} 

\vspace{.3cm}

\noindent The Sklyanin Exchange relations are the starting point to quantise  integrable field theories whilst preserving integrability by the so-called Quantum Inverse Scattering Method (QISM) \cite{Sklyanin:1980ij}. Consequently, they allow the application of  the algebraic Bethe ansatz, which is a method to derive the quantum spectrum and thus solve the theory. See e.g. \cite{Faddeev:1996iy,Korepin:1993kvr} and  the recent  notes \cite{Retore:2021wwh}, as well as a few more comments in the next section.  For us the Sklyanin relations are already sufficient in the sense that they  ensure strong classical integrability. Indeed, when taking the trace in both the tensor product arguments we find\footnote{Use that $\Tr_{12} [r_{12} , M_1 M_2] =  \left( r^{AB} \Tr T_A M \otimes \Tr T_B M - r^{AB} \Tr M T_A \otimes \Tr M T_B \right) $ as well as the cyclicity of the trace.}
\begin{equation}
 \{ \Tr_1 T_1(b, a ; z), \Tr_2 T_2 (b,a ; z' )\} = \Tr_{12} \{ T_1(b, a ; z) , T_2 (b,a ; z' ) \}  =  0 . 
\end{equation}
More generally we have,
\begin{equation}
 \{ \Tr_1 T_1(b, a ; z)^p, \Tr_2 T_2 (b,a ; z' )^q \} = \Tr_{12} \{ T_1(b, a ; z)^p , T_2 (b,a ; z' )^q \}  = 0,
\end{equation}
for all $p,q \in \mathbb{N}$ and $z, z' \in \mathbb{C}$, due to the Leibnitz rule and  the cyclicity of the trace. Hence, the classical  charges obtained from expanding the monodromy $T(z)$ are indeed in involution. For instance, when the monodromy is an analytic function near the origin we have,
\begin{equation}
 T(z) = \sum_{n=0}^\infty Q_n z^n , \qquad \partial_\tau Q_n = 0 , \quad \text{and} \quad \{Q_n , Q_m \} = 0 \ ,
\end{equation}
for all $n,m \in \mathbb{N}$.\newline
 
 \noindent $\bullet$ It turns out, however, that the integrable field theories representing  worldsheet sigma models do  not possess the Poisson bracket property of \eqref{eq:FTInvolution1}. Instead their brackets contain terms proportional to $\partial_\sigma \delta (\sigma - \sigma ' )$ and are, therefore, referred to as \textit{non-ultralocal} (the same jargon is used when higher derivatives of the delta-function are involved). In that case the QISM is not directly applicable anymore  and quantisation of the field theory becomes very challenging. In fact, generically this is still an open problem (for some progress we give a few comments at the end of section \ref{s:PCMMaillet}). 
 In some cases, however, one can for instance bootstrap the exact quantum $S$-matrix  using  extra hidden symmetries, as well as consistency properties such as unitarity and analyticity. One typically then checks this bootstrapped $S$-matrix  against loop calculations. 
 Since in these lectures we  worry only about classical integrability, it is for us sufficient when the Poisson brackets take the so-called $r/s$ Maillet form, which is another possible bracket that ensures involution of a set of charges. The Maillet brackets typically are the ones occurring for integrable worldsheet theories, they are non-ultralocal and they read,
\begin{equation} \label{eq:PoissonBracketsRSMaillet}
\begin{aligned}
\{ {\cal L}_{\sigma, 1}(\sigma , z) , {\cal L}_{\sigma, 2}(\sigma' , z') \} 
={}& \left[ r_{12}(z, z') , {\cal L}_{\sigma, 1} (\sigma, z) \right] \delta(\sigma - \sigma' )  \\&  -  \left[ r_{21}(z' , z) , {\cal L}_{\sigma, 2} ( \sigma, z') \right] \delta(\sigma - \sigma' )    \\ &  - s_{12}(z,z') \partial_\sigma \delta (\sigma - \sigma') , 
\end{aligned}
\end{equation}
with  $s_{12} (z,z') = r_{12}(z, z') + r_{21}(z', z)$. Note that in the case that the (constant) $r$-matrix is skew $r_{12} (z,z') = - r_{21} ( z' , z)$ the non-ultralocal term vanishes and we reduce to the case \eqref{eq:FTInvolution1} of above up to redefining $r \rightarrow -r$. Interestingly, the  CYBE \eqref{eq:FTcYBE} is still sufficient to ensure the Jacobi identity of the Maillet  bracket \eqref{eq:PoissonBracketsRSMaillet}, even when the $r$-matrix is non-skew. When the theory indeed has these brackets,
Maillet showed   in \cite{Maillet:1985ec,Maillet:1985ek} that by carefully 
regularising\footnote{The non-ultralocal terms in \eqref{eq:PoissonBracketsRSMaillet} produce a discontinuity in the brackets between the transport  matrices when endpoints of the integration coincide, making a naive calculation of the  Poisson brackets such as \eqref{eq:MailletBracketTM}   ill-defined. This problem has been carefully treated in \cite{Maillet:1985ek} by an infinitesimal point splitting at the endpoints and a total symmetrisation over all possibilities. The resulting regularised brackets, of which the definition depends on the number of nested Poisson brackets, are also called the Maillet brackets. Unfortunately we will not have the space-time to go through this derivation in detail, but see for instance \cite{Dorey:2006mx,Vicedo:2011zz} for a detailed exposition and review on the procedure.} 
  the Poisson brackets between the transport matrices $T(b,a ;z )$  take the form,
\begin{equation}\label{eq:MailletBracketTM}
\begin{aligned}
\{ T_1(b, a ; z) , T_2 (b,a ; z' ) \} ={}&  \left[ r(z, z') , T_1(b, a ; z) T_2 (b,a ; z' ) \right]  - T_2(b, a ; z') s(z,z') T_1 (b,a ; z ) \\&+ T_1(b, a ; z) s(z,z') T_2 (b,a ; z' ) ,
\end{aligned}
\end{equation}
where we introduced,
\begin{equation} \begin{aligned}
r(z, z' ) &\equiv \frac{ r_{21}(z' , z) - r_{12}(z, z')}{2}, \\
 s(z, z' ) &\equiv \frac{r_{21}(z' , z) + r_{12}(z, z') }{2} = \frac{s_{12}(z, z')}{2} .
\end{aligned}
\end{equation}
Hence, similar as before, taking the trace in both the tensor product arguments yields again,
\begin{equation}
\Tr_{12} \{ T_1(b, a ; z) , T_2 (b,a ; z' ) \}  = \{ \Tr_1 T_1(b, a ; z), \Tr_2 T_2 (b,a ; z' )\} = 0,
\end{equation}
and more generally,
\begin{equation}
\Tr_{12} \{ T_1(b, a ; z)^n , T_2 (b,a ; z' )^m \}  = \{ \Tr_1 T_1(b, a ; z)^n, \Tr_2 T_2 (b,a ; z' )^m \} = 0,
\end{equation}
for all $n,m \in \mathbb{N}$ and $z, z' \in \mathbb{C}$. \newline

\noindent \textbf{Short conclusion} --- With all of the above combined we can  simply conclude that we define classical integrability of a two-dimensional field theory as \textit{(i)} the possibility of recasting  the equations of motion in terms of a flat Lax connection ${\cal L}(z)$ satisfying \eqref{eq:FTLaxPair}, or equivalently \eqref{eq:LaxZeroCurv}, and \textit{(ii)} the Poisson brackets of ${\cal L}_\sigma (z)$ admitting an $r$- or $r/s$-matrix such that  the Sklyanin form \eqref{eq:FTInvolution1} or Maillet form \eqref{eq:PoissonBracketsRSMaillet} is satisfied.\footnote{Another definition of a classical integrable field theory that is sometimes taken in the literature is to  have  an infinite set of \textit{local} charges in involution, without necessarily referring to Lax connections, monodromy matrices and Maillet brackets. This definition is, however, less common in the community of integrable string sigma models.}
Let us finally also recall that when speaking of the ``classical integrable structure" one usually refers to the data of the Lax connection ${\cal L}(z)$, the wave function $\Psi(z)$ and the monodromy or transport matrix $T(z)$. As we will comment in the next subsection, the $r$-matrix on the other hand plays an important role in the ``quantum integrable structure". Hence, for classical integrability one usually gets already far with the weak version of its definition.

%{\color{red}Finally, .... comments on "integrable structure" and r-matrix stuff more for quantum integrable structure}

\subsection{Comments on quantum integrability and properties of $r$-matrices} \label{s:qint}

Although these lectures only discuss classical integrability, it would be odd not to mention some of the hallmarks of the quantum integrable regime. In this section we  will broadly introduce the main ones. For  more details we refer e.g.~to the following  lecture notes \cite{Retore:2021wwh,Bombardelli:2016rwb}.

\begin{figure}[H]
\centering
\includegraphics[scale=.2]{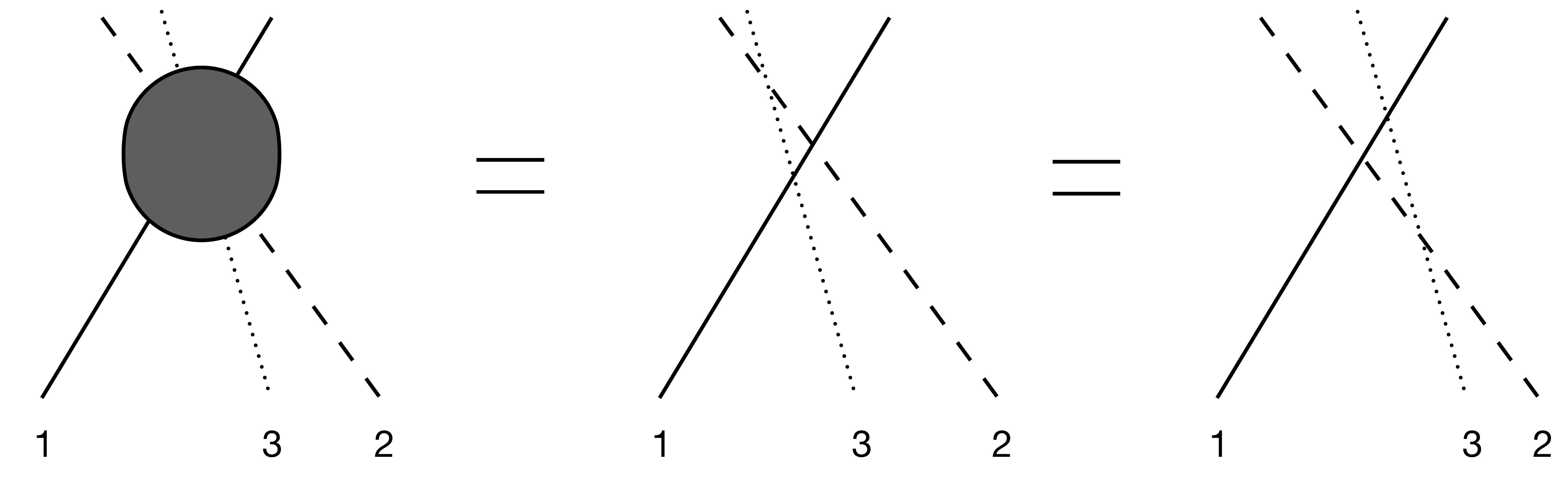}
\caption{A $3 \rightarrow 3$ scattering process factorised in three $2\rightarrow 2$ scattering processes of which the order does not matter: scattering particles $2,3$, then $1,3$ and finally $1,2$ yields the same result as scattering particles $1,2$, then $1,3$ and finally $2,3$. Time flows from left to right.}
\label{f:qybe}
\end{figure}

The property of integrability in the field theory transfers to the quantum regime when demanding that the classical  conserved charges and their corresponding symmetries do not suffer from anomalies.\footnote{See for instance \cite{Goldschmidt:1980wq} for a counting argument on quantum anomaly terms that may spoil the classical conservation laws of  integrable symmetric space sigma models.} The main hallmark of a quantum integrable field theory is that the large number of symmetries constrain the quantum $S$-matrix in an  extremely severe way:\footnote{The properties that follow are particularly implied by the existence of  \textit{local} higher spin conserved charges. Other symmetries, for example following from non-local charges, may constrain the $S$-matrix even further.}  \textit{(i)} the theory does not exhibit particle production or annihilation, \textit{(ii)}  individual particle momenta are conserved $\{p'_1 , \ldots , p'_n\} = \{p_1 , \ldots , p_n\}$, and    \textit{(iii)} $n\rightarrow n$ scattering processes are all factorised in $2\rightarrow 2$  processes   of which the order does not matter \cite{Zamolodchikov:1978xm,Parke:1980ki} (see also the notes \cite{Dorey:1996gd,Bombardelli:2016scq}). The latter statement is captured by the \textit{Quantum Yang-Baxter} equation (QYBE), which is usually expressed through the so-called
 quantum $R$-matrix,
\begin{equation}
  R_{23} R_{13} R_{12}  = R_{12} R_{13} R_{23} \ .
\end{equation}
For the $S$-matrix this equation translates to the equality of the scattering processes depicted in figure \ref{f:qybe}.

\noindent  It is generally assumed that the quantum $R$-matrix depends on an additional parameter $\hbar$. Expanding it as 
%\footnote{Similarly, the monodromy matrix can be extended to the quantum monodromy matrix
%\begin{equation}
%\hat{T}(z) =  T(z) + {\cal O}(\hbar) ,
%\end{equation}
%and the quantisation of the Sklyanin Exchange relations form the famous $RTT$-relations,
%\begin{equation}
%R_{12}(z,z') \hat{T}_1(z) \hat{T}_2(z') = \hat{T}_2(z') \hat{T}_1(z) R_{12}(z,z') ,
%\end{equation}
%which underlie the QISM and the algebraic Bethe ansatz. Note that, again, $\hat{T}$ is a special $R$-matrix. They imply that the trace of quantum monodromy matrices at different values of the spectral parameter commute , similar to the classical case,  and thus that they have a common basis of eigenvectors. {\color{red} merge this in the text}
%}
\begin{equation}
R_{ij} = 1 \otimes 1 + i \hbar r_{ij} + {\cal O}(\hbar^2) ,
\end{equation}
 one can show that   at second order in $\hbar$ the QYBE for $R_{ij}$ produces precisely the CYBE for $r_{ij}$. Interestingly the quantum $R$-matrix can thus be  related to a large class of classical $r$-matrices.\footnote{Recall that the CYBE is a consistency equation for large classes of $r$-matrices, but not for the most general ones.} The quantum $S$-matrix is  a particular (exact) $R$-matrix which  is also unitary, crossing symmetric and respects the symmetries of the theory.\newline

Another particular quantum $R$-matrix is the quantum monodromy, 
\begin{equation}
\hat{T}(z) =  T(z) + {\cal O}(\hbar) .
\end{equation}
which must satisfy the quantum version of the  Sklyanin Exchange relations known as the famous RTT relations,
\begin{equation} \label{eq:RTT}
R_{12}(z,z') \hat{T}_1(z) \hat{T}_2(z') = \hat{T}_2(z') \hat{T}_1(z) R_{12}(z,z') .
\end{equation}
The RTT relations particularly underlie quantum methods such as the QISM and the algebraic Bethe ansatz. When we add an auxiliary space indicated by the index $3$   as\footnote{At the end of the day one needs to trace out the auxiliary space.}
\begin{equation}
R_{12}(z,z') \hat{T}_{1{\color{red} 3}}(z) \hat{T}_{2{\color{red} 3}}(z') = \hat{T}_{2{\color{red} 3}}(z') \hat{T}_{1{\color{red} 3}}(z) R_{12}(z,z') ,
\end{equation}
and taking $R=\hat{T}$,
we see indeed  that $\hat{T}$ satisfies the QYBE and thus that it is a special $R$-matrix. Note, however, that other solutions to the RTT relations may exist.  The RTT relations \eqref{eq:RTT} imply that the trace of quantum monodromy matrices at different values of the spectral parameter commute
\begin{equation}
\left[ \Tr \hat{T}_1 (z) , \Tr \hat{T}_2(z') \right] = 0 \ .
\end{equation}
 This is similar to the classical case,  and it shows that there exists a common basis of eigenvectors. For more details, as well as an introduction to the QISM and the algebraic Bethe ansatz, see the pedagogical review \cite{Faddeev:1996iy}.\\

The main take-home message of this section is the following. Imagine that one wants to reverse the logic, and not check integrability  for a particular Lagrangian or Hamiltonian theory, but rather wants to find (or even classify) consistent  integrable structures,  in particular $r$-matrices, and construct from them the charges and the $S$-matrix (see e.g. \cite{Retore:2021wwh,Faddeev:1996iy}). We have seen here
that properties of large classes of  classical $r$-matrices, namely those that solve the  CYBE \eqref{eq:FTcYBE}, will strongly restrict the quantum integrable structure  through the QYBE. Such solutions indeed govern the scattering of the field theory, as well as  the quantum conserved charges and symmetries  generated from the (quantum) monodromy matrix. Interestingly, the QYBE relates  to intriguing mathematical frameworks like quantum groups, see e.g.~\cite{Loebbert:2016cdm}. Let us very briefly discuss some important results from \cite{BelDri82,Belavin1998TriangleEA,BelDri83} in this respect.\\

\noindent \textbf{Belavin-Drinfeld theorems} --- Assume that $\mathfrak{g}$ is a simple Lie algebra with non-vanishing Coxeter number, and that  the $r$-matrix satisfies the CYBE \eqref{eq:FTcYBE}.  When $r(z,z')$ has a simple pole at $z-z'$ with a residue proportional to $C_{12}$, then Belavin and Drinfeld showed that $r$ can be transformed to a difference form $r(z-z')$ \cite{BelDri83}. Furthermore, in that case $r_{12} (z-z') = - r_{21}(z'-z)$ and the matrix can be extended to the full complex plane, on which all of the poles are simple and form a lattice $\Gamma$. Depending on the dimension $d$ of $\Gamma$, the $r$-matrix is ``elliptic" ($d=2$), ``trigonometric" ($d=1$) or ``rational" ($d=0$, $\Gamma = \{0\}$), meaning it depends respectively on elliptic, trigonometric and rational functions only. The Belavin-Drinfeld theorems thus provide an important classification of classical $r$-matrices under the above assumptions. As argued above, they therefore also severely constrain the possible quantisations and  the associated quantum integrable theories.  In particular, the quantum theories will have an underlying infinite-dimensional quantum group symmetry known as the ``elliptic quantum group" ($d=2$), ``quantum affine algebras" ($d=1$) or ``Yangian algebras" ($d=0$).\footnote{\color{black}The Belavin-Drinfeld results  indicate that there is an intimate relationship between the poles of the $r$-matrix and the integrable field theory. For recent interesting works exploring such relations from other viewpoints see for instance \cite{Vicedo:2017cge,Vicedo:2019dej,Lacroix:2020flf}.}  \newline

    \begin{exercise} \label{ex:rmatrix}
The typical example of a rational $r$-matrix is the famous \textit{Yang's r-matrix},
\begin{equation}
r_{12}(z , z' ) = \kappa \frac{C_{12}}{z-z'} ,
\end{equation}
for $\kappa$ some proportionality constant. It captures the non-linear Schr\"odinger model. Proof that Yang's r-matrix solves the CYBE \eqref{eq:FTcYBE}. Show that in fact this $r$-matrix enjoys the freedom of a multiplicative function in $z$. In other words,
\begin{equation}
r_{12}(z , z' ) = \kappa \frac{C_{12}}{z-z'} \phi^{-1}(z),
\end{equation}
for some arbitrary $\phi(z)$  (called the twist function)  will also solve the CYBE. This $r$-matrix captures the Principal Chiral Model (PCM) and many of its integrable deformations, for which the twist function gets deformed.
%These two classes of $r$-matrices  connect to Yangians and ``deformed" Yangians, or quantum groups, see e.g.~\cite{Loebbert:2016cdm}.
    \end{exercise}

\subsection{A non-exhaustive list of examples}

In this section we give a non-exhaustive list of  well-known two-dimensional integrable field theories.

\begin{itemize}
\item \textit{The Korteweg-De Vries equation.} This model describes waves $\phi$ in shallow water through the following  non-linear PDE,
\begin{equation}
\partial_\tau \phi + \partial_\sigma^3 \phi - 6 \phi \partial_\sigma \phi = 0  .
\end{equation}
 It is the prototypical example of an exactly solvable model which has solitonic solutions, i.e.~wave forms which maintain their shape and travel at a constant speed. This can be understood as a consequence of the extra conservation laws coming from the tower of conserved charges. For an illustration of solitary waves see figure \ref{fig:solitons}. 
 
\begin{figure}[h]
\centering
\begin{subfigure}{.5\textwidth}
  \centering
  \includegraphics[width=.8\linewidth]{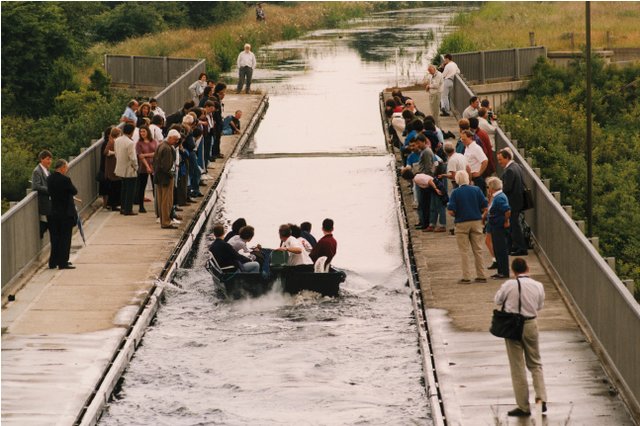}
  \caption{Recreation of a solitary wave in the Union canal (Edinburgh, 1995) discovered by John Scott Russell in 1834 (Image: Mathematics, Heriot-Watt University).}
  \label{fig:solsub1}
\end{subfigure}%
\begin{subfigure}{.5\textwidth}
  \centering
  \includegraphics[width=.8\linewidth]{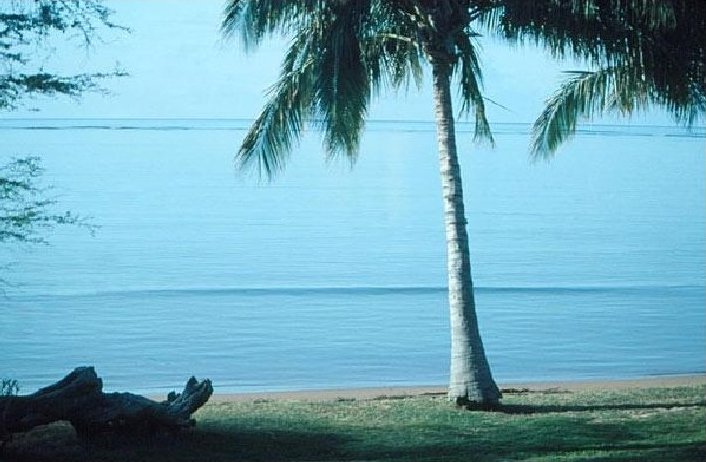}
   \caption{A solitary wave travelling to Hawa\"i (Image: Robert I.~Odom, University of Washington).\\ }
  \label{fig:solsub2}
\end{subfigure}
\caption{Illustrations of solitary waves.}
\label{fig:solitons}
\end{figure}

\item \textit{The Sine-Gordon Model}. Also the Sine-Gordon equation for a real scalar $\phi$ has solitonic solutions,
\begin{equation}
\partial_\tau^2 \phi - \partial_\sigma^2 \phi = -\frac{8 m^2}{\beta} \sin (2\beta\phi) .
\end{equation}
This model has numerous applications ranging from condensed matter to differential geometry and string theory. It is an example of an \textit{Affine Toda field theory} which are theories with an underlying affine Kac-Moody algebra.
\item \textit{The Non-Linear Schr\"odinger Model}. This  classical model describes a complex field $\phi$ with dynamics controlled by an equation which is similar to the usual (quantum) Schr\"odinger equation for models with a non-linear potential proportional to the modulus squared of the wave function,
\begin{equation}
i \partial_\tau \psi = \partial_\sigma^2 \psi + 2\kappa | \psi|^2 \psi . 
\end{equation}
Depending on the sign of $\kappa$ it describes solitons with different properties (``dark" or ``bright" solitons). The model has  a wide range of applications, most notably in condensed matter physics in the mean field regime.
\item \textit{The Principal Chiral Model (PCM).} We will describe this model in a lot of detail in the next section.
\item \ldots 
\end{itemize}

\newpage

\section{Principal Chiral Model}\label{s:PCM}

To illustrate the  concepts of the previous chapter, we will consider here a distinguished example of an integrable non-linear sigma-model, known as the Principal Chiral Model (PCM). The interest in the  PCM started in the mid '70s notably due to Belavin, L\"uscher, Migdal, Mikhailov, Pohlmeyer, Polyakov and Zakharov  \cite{Polyakov:1975rr,Polyakov:1975yp,Migdal:1975zg,Pohlmeyer:1975nb,Luscher:1977rq,Luscher:1977uq,Zakharov:1973pp}, who were stimulated by the pioneering work of Gross, Politzer and Wilczek \cite{Gross:1973id,Politzer:1973fx}. They  were particularly motivated by the fact that the PCM can be seen as an \textit{integrable} two-dimensional prototype for four-dimensional Yang-Mills theory. In fact, the PCM theory  is asymptotically free in the UV, while it  becomes strongly coupled and dynamically generates a mass gap in the deep IR. The power of its (quantum) integrability has been  illustrated notably in \cite{Hasenfratz:1990zz,Balog:1992cm}  where  an exact expression for its mass gap   has been obtained. In string theory, the PCM is interesting since the interpretation of this model as a string worldsheet theory has been key in the construction of explicit AdS/CFT examples. In particular, although the PCM does not describe a conformally invariant worldsheet, it can be present in a subsector of a consistent curved  background as a bosonic truncation in which fermions have been set to zero.
For example, the PCM describes the $\mathrm{S}^3$ of the $\mathrm{AdS}_3\times \mathrm{S}^3\times \mathrm{T}^4$ type IIB superstring background supported by RR fields. Another (generalised) example, is the well-known $\mathrm{AdS}_5 \times \mathrm{S}^5$ background of which the worldsheet theory is a generalisation of a {\color{black}PCM based on the $PSU(2,2 |4)$ superalgebra in which a further $Sp(1,1) \times Sp(2)$ subgroup has been gauged away \cite{Metsaev:1998it}.} And in this way the list goes on.

In this chapter, we will introduce the PCM action functional and we will proof its strong classical integrability. We represent the equations of motion with a flat Lax connection, discuss the conserved charges that generate  hidden symmetry algebras, and give their Poisson algebra structure in terms of the Maillet bracket. However, before doing so  we will briefly introduce a generic two-dimensional non-linear sigma model. 

\subsection{Intermezzo: non-linear sigma-models for string theory} \label{s:sigmamodels}

In this section we give some of the salient features of non-linear sigma-models (within the interpretation as a string worldsheet) to ease the introduction of the Principal Chiral Model in the next section. At the same time this allows us to lay out our conventions. Readers familiar with non-linear sigma-models for string theory may easily skip this section.\\

Two-dimensional non-linear sigma-models are particularly interesting for us since they appear as  string worldsheet theories that give rise to non-trivial curved backgrounds $M$, i.e.~backgrounds carrying a curved  spacetime metric $G$ and a Kalb-Ramond two-form $B$. Let us, however, consider a string propagating in a flat spacetime first. While doing so, it is sweeping out a $(1 + 1)$-dimensional surface $\Sigma$ called the worldsheet, which we parametrise by a time $\tau$ and a spatial $\sigma$ coordinate as $\sigma^\alpha = (\tau, \sigma) $ with $\alpha = 1,2$. The string’s dynamics is encoded in the so-called Polyakov action,
\begin{equation}
S(x,g) = \frac{1}{4\pi\alpha'} \int_\Sigma d^2\sigma \sqrt{-g} g^{\alpha\beta} \partial_\alpha x^\mu \eta_{\mu\nu} \partial_\beta x^\nu ,
\end{equation}
which can be derived from a minimal action principle; that is, by minimising its worldsheet area $\text{Vol}(\Sigma)$ for which $\alpha'$ measures the tension $T =(2\pi\alpha')^{-1}$. The object $g_{\alpha\beta}$ is a (dynamical)  metric on the worldsheet, $\eta_{\mu\nu}$ is the flat spacetime metric, and $x^\mu = x^\mu(\sigma^\alpha)$, with $\mu = 0, \ldots, D-1$, are the spacetime coordinates labelling the position of the string in spacetime.   Hence, these coordinates can be viewed as maps from the worldsheet $\Sigma$ to the flat spacetime $\mathbb{R}^{1,D-1}$,
\begin{equation}
x^\mu \ : \ \Sigma \rightarrow \mathbb{R}^{1,D-1} \ : \ \sigma^\alpha \mapsto x^\mu (\sigma^\alpha) .
\end{equation}
Alternatively, we could   abandon  the string and spacetime picture, and view the Polyakov action as a $(1+1)$-dimensional field theory with a dynamical metric $g$, where the maps $x^\mu$ are interpreted as $D$ free scalar bosonic fields living on the worldsheet. Such a theory of maps, from a \textit{base space} $\Sigma$ to a \textit{target space} $M$, and in which we can take both point of views, is called a sigma model (albeit here to a flat target space). 

The Polyakov action enjoys a number of symmetries. Apart from the global Poincar\'e spacetime invariance, the worldsheet theory is invariant under two gauge symmetries: diffeomorphisms and Weyl rescalings, where the latter is particularly special to two-dimensional dynamical theories. 
One often fixes these gauge symmetries by setting the dynamical worldsheet metric $g_{\alpha\beta}$ to be a flat Minkowski metric $\eta_{\alpha\beta}$, for which we take the following convention:
\begin{equation} \label{eq:confgauge}
 \eta_{\alpha\beta} = \mathrm{diag}(+1 , -1 ).
\end{equation}
This is known as the conformal gauge. It introduces mass-shell constraints in our system, the Virasoro constraints, which can be formulated in terms of the energy momentum tensor as,
\begin{equation}
T_{\alpha\beta} \approx 0 , \quad \text{where} \quad  T_{\alpha\beta} = \frac{4 \pi  }{\sqrt{-g}} \left. \frac{\delta S}{\delta g^{\alpha\beta}} \right\vert_{g_{\alpha\beta} = \eta_{\alpha\beta}} ,
\end{equation}
in which the overall factor is a matter of convention. Important in this discussion  is that  choosing the conformal gauge has not completely fixed the gauge. There are residual gauge symmetries, diffeomorphisms that  can be compensated by a Weyl rescaling,\footnote{Such diffeomorphisms are also called conformal reparametrisations; they change the coordinates only holomorphically.} which preserve the choice of the worldsheet metric. These diffeomorphisms are   conformal symmetries and thus the gauge-fixed action, $S(x) = S(x, g_{\alpha\beta} = \eta_{\alpha\beta})$,
describes a two-dimensional conformal field theory of free scalar fields $x^\mu$ on a dynamical worldsheet.
For more details on, e.g., the derivation of the Polyakov action and its symmetries, we refer the reader to the following excellent introductory notes to string theory  \cite{Tong:2009np}. \\

After quantisation one can derive the spectrum from the mass-shell Virasoro constraints to find (besides tachyonic excitations) three fundamental massless excitations: a spin two graviton $G_{\mu\nu}(x)$, the antisymmetric $B$-field $B_{\mu\nu}(x)$ and the dilaton $\Phi(x)$. Considering then a collection of  massless excitations of  flat space strings, one can produce a curved background in which one can consider the propagation of another string. In particular, one constructs in this way a curved spacetime with a curved metric $G_{\mu\nu}(x)$ which is truly built out of quantised graviton states (or vertex operators).  The corresponding string action  then is
\begin{equation}
S(x,g) = \frac{1}{4\pi\alpha'} \int_\Sigma d^2\sigma \sqrt{-g} \left( g^{\alpha\beta} \partial_\alpha x^\mu G_{\mu\nu}(x) \partial_\beta x^\nu +\epsilon^{\alpha\beta} \partial_\alpha x^\mu B_{\mu\nu} (x) \partial_\beta x^\nu + \alpha' \Phi (x) R \right), 
\end{equation}
where $R$ is the worldsheet Ricci scalar and $\epsilon_{\alpha\beta}$ is the Levi-Civita tensor   which we normalise as $\sqrt{-g}\epsilon_{01} =  1$.\footnote{In these conventions one has after conformal gauge for a general $r$-form $\omega$ on $r$-dimensional extensions of $\Sigma$ with a mostly-minus metric that $\omega_{\alpha_1 \cdots \alpha_r} d\sigma^{\alpha_1} \wedge \cdots \wedge d\sigma^{\alpha_r} = (-)^{r+1} \omega_{\alpha_1 \cdots \alpha_r}\epsilon^{\alpha_1 \cdots \alpha_r} d\sigma^0 \wedge \cdots d\sigma^r$ and hence the $B$-term can be written as $S_B(x) = -\frac{1}{2\pi\alpha'}\int_\Sigma B$.}
This theory is still a non-linear sigma model, now from the base space $\Sigma$ to a curved target space $M$ equipped with a curved metric $G_{\mu\nu}(x)$.\footnote{In the literature, the term (non-linear) sigma model is mostly used for the action with only the curved spacetime metric non-vanishing (and thus no $B$-field and dilaton excitations) and with the worldsheet metric $\eta_{\alpha\beta}$ non-dynamical. The classical conformal symmetry of these theories is in general anomalous, which is not problematic because the conformal symmetry is not a gauge symmetry for non-dynamical $\eta_{\alpha\beta}$. This is in fact the situation for  the Principal Chiral Model. At present the term sigma model is often used interchangeably in the string community to study strings in the three background fields $G,B,\Phi$. The meaning should be clear from the context.}
In the rest of these notes we will ignore the $B$-field and dilaton excitations and we will always work in the conformal gauge $g_{\alpha\beta} = \eta_{\alpha\beta}$. The classical  action, in light-cone coordinates $\sigma^\pm =\tau \pm \sigma$, then takes the form
\begin{equation} \label{eq:GenActionLightCone}
S(x) = \frac{1}{\pi \alpha'} \int_\Sigma d^2 \sigma \  \partial_+ x^\mu G_{\mu\nu}(x) \partial_- x^\nu .
\end{equation}
In addition, we will mainly remain in the classical regime, and thus we will not need to worry too much about the mass-shell Virasoro constraints. Let us, however, for completeness state them explicitly here
\begin{equation}\label{eq:EMGen}
T_{\alpha\beta} = \frac{1}{\alpha'} \left( \partial_\alpha x^\mu G_{\mu\nu}(x) \partial_\beta x^\nu - \frac{1}{2} \eta_{\alpha\beta} \eta^{\gamma\delta} \partial_\gamma x^\mu G_{\mu\nu}(x) \partial_\delta x^\nu \right) \approx 0 , \\
\end{equation}
and thus in light-cone coordinates $T_{+-} = T_{-+} = 0 $ and $ T_{\pm \pm } = \frac{1}{\alpha'} \partial_\pm x^\mu G_{\mu\nu}(x) \partial_\pm x^\nu \approx 0$. Furthermore notice also that the worldsheet sigma model  itself is now   interacting  and in fact must correspond to a two-dimensional CFT of $D$  interacting scalar fields $x^\mu$.  There are an infinite number of coupling parameters, as can for instance be observed by  Taylor expanding the scalars $x^\mu(\sigma)$ around a classical solution $x^\mu_{\text{\tiny cl}} $ in the action \eqref{eq:GenActionLightCone}. Each derivative of the metric can then be thought of as  coupling constants for interactions in the fluctuations, and  all of them are  accompanied with a factor $\sqrt{\alpha'}$ so that the worldsheet perturbation series is an $\alpha'$-expansion with $\alpha'$  the loop counting parameter. This perturbative expansion gives rise to a renormalisation group flow of the coupling constants, which are famously captured by the  $\beta$-functions. To remain a CFT at the quantum level, these $\beta$-functions should vanish. When only the metric coupling is present (as it will be for the Principal Chiral Model) they read \cite{Friedan:1980jf}
\begin{equation}
\mu \frac{d}{d\mu} G_{\mu\nu} = \alpha'  {\cal R}_{\mu\nu} + {\cal O}(\alpha'^2) \ ,
\end{equation}
with ${\cal R}_{\mu\nu}$ the Ricci tensor. Here we see how a worldsheet symmetry can severely constrain the dynamics of the target space background and geometry: requiring $\mu \frac{d}{d\mu} G_{\mu\nu} $ to vanish forces the target space to satisfy Einstein's equations in the absence of sources. In addition, we find a very easy and generic way to compute $\beta$-functions: we simply need the target space geometry, in this case the target space metric, of the sigma model. The same is in fact true when non-trivial $B$- and $\Phi$-couplings are turned on, in which case the target space geometry must be a (bosonic) supergravity.
The Principal Chiral Model in general is not, however,   Ricci-flat   and thus the above $\beta$-function will not vanish. Instead, its target space metric changes with scale---at high energies the geometry flattens out while at low energies it becomes highly curved---which relates to an important concept in differential geometry known as Ricci flow. The $\beta$-functions of  generic sigma models (not necessarily CFTs) thus capture interesting physics, but if one insists on a string theory application, these sigma models can be viewed as (bosonic) truncations that should be “improved" in such a way that the geometry ensures conformal invariance on the worldsheet.\footnote{The PCM model, for instance, can be promoted to a CFT by adding a Wess-Zumino term which allows to tune the overall coupling constants in such a way that the worldsheet is, in fact, \textit{exactly} conformal invariant \cite{Witten:1983ar}. The corresponding model is famously known as the Wess-Zumino-Witten model.}

For more details, let us again refer to \cite{Tong:2009np} but also the classic TASI notes \cite{Callan:1989nz}.

\subsection{PCM action, equations of motion and canonical formulation} \label{s:PCMaction}

The PCM is a  non-linear sigma model of maps  from a two-dimensional Riemann surface (or worldsheet) $\Sigma$ to a Lie group {manifold} $G$.\footnote{For example, for $G=SU(2)$ the target space is the three-sphere $S^3$.} It will therefore be more convenient to denote the maps or  field configurations  of the PCM theory by $g(\tau,\sigma) \in G$, rather than the maps $x^\mu$ used in the previous section.  In particular, 
\begin{equation}
g(\tau, \sigma) \;\; : \;\; \Sigma  \ \rightarrow \  G  \ .
\end{equation}
They are assumed to satisfy either periodic boundary conditions,
\begin{equation} \label{eq:periodicboundary}
g(\tau, \sigma) = g(\tau, \sigma + 2\pi) ,
\end{equation}
or asymptotic  boundary conditions,
\begin{equation} \label{eq:asymptoticboundary}
g(\tau , \sigma \rightarrow \pm \infty ) = \text{constant} .
\end{equation}
The former are considered for string sigma models describing propagating closed strings. The latter are however more convenient to  discuss the classical integrable structure and, in particular, some of  the hidden symmetries as we will see below.  \\
\indent The PCM is defined by the action,
\begin{equation} \label{eq:PCMaction}
S_{\text{\tiny PCM} , \kappa^2}(g)  =  \frac{\kappa^2}{4\pi} \int_\Sigma d\sigma d\tau \eta^{\alpha\beta}\langle \partial_\alpha g^{-1} , \partial_\beta g \rangle = - \frac{\kappa^2}{4\pi} \int_\Sigma d\sigma d\tau \langle g^{-1} \partial_\alpha g , \eta^{\alpha\beta} g^{-1} \partial_\beta g \rangle ,
\end{equation}
with $\kappa$ an (inverse) coupling constant, $\eta_{\alpha\beta}$ the worldsheet metric in conformal gauge as in \eqref{eq:confgauge}, and $\langle \cdot, \cdot \rangle$ a non-degenerate symmetric and ad-invariant bilinear form on the Lie algebra $\mathfrak{g}$ of $G$.
Let us at this stage lay out some of our group and algebra notations and conventions.
 We assume that $G$ is real and that we have a matrix realisation of $G$ and $\mathfrak{g}$.  For the bilinear form we then take  the one induced by the trace of products of matrices. Namely, considering the basis of generators $T_A$, $A = 1, \ldots , \dim \mathfrak{g}$, then $\langle T_A , T_B \rangle = \Tr (T_A T_B) \equiv \eta_{AB}$. Ad-invariance then simply means that $\Tr (\text{Ad}_g(T_A) \text{Ad}_g(T_B)) = \Tr (T_A T_B) $ with $\text{Ad}_g (X) = g X g^{-1}$ for $X\in \mathfrak{g}$. Lastly, our conventions are to expand Lie algebra elements as $X = -i X^A T_A $ and have the commutation relations $[T_A , T_B] = i F_{AB}{}^C T_C$ with the components $X^A$ and structure constants $F_{AB}{}^C$ real.

The PCM theory has a  number of interesting properties. It enjoys a global (internal) $G_L \times G_R$ invariance which acts independently as,
\begin{equation} \label{eq:GLGRPCM}
g (\tau, \sigma) \rightarrow g_L \,  g (\tau, \sigma)  ,\quad \text{and} \quad g (\tau, \sigma)  \rightarrow g (\tau, \sigma)\,  g_R , 
\end{equation}
for any constant element $g_L , g_R \in G$. The corresponding conserved  currents are related and given by,
\begin{equation} \label{eq:PCMcurrents}
R_\alpha =   \partial_\alpha g g^{-1}, \quad \text{and} \quad L_\alpha = g^{-1} R_\alpha g = g^{-1} \partial_\alpha g,
\end{equation}
 respectively. This can be seen directly from the classical equations of motion which  are in fact equivalent to their conservation laws. When varying the action \eqref{eq:PCMaction} with respect to the fields $g$ subjected to the periodic boundary conditions \eqref{eq:periodicboundary} or asymptotic boundary conditions \eqref{eq:asymptoticboundary} we find,
\begin{equation}\label{eq:PCMEOM0}
\begin{aligned}
\delta S_{\text{\tiny PCM} , \kappa^2} &= \frac{\kappa^2}{2\pi} \int_\Sigma d\sigma d\tau \langle g^{-1} \delta g , \eta^{\alpha\beta} \partial_\alpha (g^{-1} \partial_\beta g ) \rangle, \\
&= \frac{\kappa^2}{2\pi} \int_\Sigma d\sigma d\tau \langle  \delta g g^{-1} , \eta^{\alpha\beta} \partial_\alpha ( \partial_\beta g g^{-1} ) \rangle, 
\end{aligned}
\end{equation}
where we have used the ad-invariance of the bilinear form in the second line. 
\vspace{.1cm}

    \begin{exercise}
Show  eq.~\eqref{eq:PCMEOM0}.
    \end{exercise}
       
\vspace{.2cm}

\noindent Now, we can write an infinitesimal variation of the $G_L \times G_R$ transformation as,
\begin{equation}
\delta_L g  = X_L g , \qquad \delta_R g = g X_R ,
\end{equation}
with $X_L$ and $X_R \in \mathfrak{g}$. In this way, we see  explicitly  that the \textit{left}-invariant Noether current $L_\alpha$ corresponds to the \textit{right} group translations $G_R$, while the  \textit{right}-invariant Noether current $R_\alpha$ corresponds to the \textit{left} group translations $G_L$ (yes, this is unfortunately confusing). As advertised, their conservations laws are indeed precisely the classical equations of motion,
\begin{equation} \label{eq:PCMEOM1}
\partial_\alpha L^\alpha = \partial_\alpha (g^{-1} \partial^\alpha g ) = 0 , 
\end{equation}
and equivalently $\partial_\alpha R^\alpha = 0$. The corresponding (local) Noether charges are respectively:
\begin{equation} \label{eq:PCMfirstcharges}
\begin{aligned}
Q^{(0)}_R  = \int_\Sigma d \sigma \, L_\tau , \qquad
Q^{(0)}_L = \int_\Sigma d\sigma \, R_\tau ,
\end{aligned}
\end{equation}
in which the path along $\sigma$ is at a fixed time-slice $\tau$. The index $(0)$ will become meaningful when we discuss hidden symmetries later.

In a differential geometry language the currents $L_\alpha$ and $R_\alpha$ are respectively known as the left- and right-invariant Maurer-Cartan one-forms taking values in the Lie algebra $ \mathfrak{g}$.\footnote{To be more precise, the Maurer-Cartan one-forms are $L=g^{-1}dg$ and $R=dg g^{-1} \in \mathfrak{g} \times \Omega^1 (\Sigma) $. They can be expanded on the worldsheet as $L=g^{-1} d g  = g^{-1}\partial_\alpha g \ d\sigma^\alpha$ with $L_\alpha =g^{-1}\partial_\alpha g \in T_e G \simeq \mathfrak{g}$ and similarly for $R =R_\alpha d\sigma^\alpha$. }
 They satisfy  the famous Maurer-Cartan identity (also called the flatness condition\footnote{The Maurer-Cartan one-forms define a co-frame  on the Lie group manifold to the frame $\{T_A \}$ spanned by the Lie algebra directions. From Cartan's formulation of general relativity one can easily infer that the Maurer-Cartan identity shows the existence of a torsionful but curvature-free connection in which case the Lie group manifold is \textit{flat}. This is known as parallelizability of Lie group manifolds. The corresponding sigma model is the WZW model \cite{Witten:1983ar} which is an appropriately normalised extension of the PCM supporting the existence of  torsion. The PCM model itself, on the other hand, corresponds to a different connection allowed by the Maurer-Cartan identity; that is, a torsion-free but curvature-full connection.}),
\begin{equation} \label{eq:MCleft}
\partial_\alpha L_\beta - \partial_\beta L_\alpha + \left[ L_\alpha , L_\beta \right] = 0  ,
\end{equation}
or equivalently,
\begin{equation} \label{eq:MCright}
\partial_\alpha R_\beta - \partial_\beta R_\alpha - \left[ R_\alpha , R_\beta \right] = 0 .
\end{equation}
From now on we focus mostly on the left currents. An important observation is that using the Maurer-Cartan identity we can  completely formulate the PCM in terms of the currents rather than the  fields $g$. Indeed, we can replace the single second-order equation of motion \eqref{eq:PCMEOM1} in terms of $g$ by two first-order equations of motion in terms of $L = g^{-1} dg$, eq.~\eqref{eq:PCMEOM1} and eq.~\eqref{eq:MCleft}, which in form language\footnote{In 1+1 dimensions the Hodge star $\star$ acting on one-forms $a,b$ has the properties $\star^2 a=a$ and $\star a\wedge b = - a\wedge\star b$. In particular, on the worldsheet coordinates we have  $\star\mathrm{d}\tau = \mathrm{d}\sigma$ and $\star\mathrm{d}\sigma = \mathrm{d}\tau$ or $\star \mathrm{d}\sigma^\pm = \pm \mathrm{d}\sigma^\pm$, with $\sigma^\pm =  \tau \pm \sigma$ the light-cone coordinates.} are,
\begin{equation} \label{eq:PCMEOM2}
\begin{aligned}
&\mathrm{d} \star L = 0 , \\
&\mathrm{d} L + L \wedge L = 0
\end{aligned}
\end{equation}
Doing so, we have simultaneously replaced the point of view of the field $g$ as fundamental to the point of view of the currents $L$ as  fundamental. To travel between the two pictures one uses the definition (or pure-gauge condition for the flatness condition) given in eq.~\eqref{eq:PCMcurrents}. Proving (weak) classical integrability will be very simple in the latter picture. \\

In the picture of the currents the PCM action simply becomes
\begin{equation} \label{eq:actionPCML}
S_{\text{\tiny PCM}, \kappa^2} (L) = - \frac{\kappa^2}{\pi} \int_\Sigma d^2 \sigma \langle L_+ , L_- \rangle  .
\end{equation}
Let us briefly introduce particular local coordinates $x^\mu$, $\mu =1, \ldots , \text{dim }(G)$, on $M=G$ as $g = g(x^\mu(\sigma^\alpha))$. We  can then expand the Maurer-Cartan form also as $L = L_\alpha d\sigma^\alpha = L_\mu dx^\mu$ using the pull-back map $x^\mu(\sigma^\alpha)$ between worldsheet and target space, i.e.~simply $dx^\mu = \partial_\alpha x^\mu d\sigma^\alpha$. Substituting  furthermore the expansion of $L$ in the Lie algebra generators as $L = -i L^A_\mu T_A dx^\mu$ in the action \eqref{eq:actionPCML} and comparing the result with \eqref{eq:GenActionLightCone} we find that the target space carries the (Lie group) metric
\begin{equation}
G_{\mu\nu}(x) = L_\mu^A \eta_{AB} L_\nu^B \ ,
\end{equation}
with $\eta_{AB} = \langle T_A , T_B \rangle$ the constant Lie algebra metric introduced above.
For generic Lie groups, this metric will be curved and thus has a non-vanishing Ricci tensor. An exception is $G$ abelian, in which case $L_\mu^A = \delta_\mu^A$, and thus the PCM reduces to a flat Polyakov-like theory. For completeness,  let us write down some final important objects of the PCM in the picture of the currents. For instance the classical energy momentum tensor encoding worldsheet symmetries  can be found to be,\footnote{\label{f:EMtensor}After coupling the PCM action  \eqref{eq:PCMaction} to a dynamical two-dimensional background  metric $g_{\alpha\beta}$, we take the following convention to define  the energy momentum tensor,
\begin{equation}
 T_{\alpha\beta} = - \frac{4 \pi}{\sqrt{-g}}  \left. \frac{\delta S}{\delta g^{\alpha\beta}} \right\vert_{g_{\alpha\beta} = \eta_{\alpha\beta}}.
 \end{equation} }
 \begin{equation}
  T_{\alpha\beta} = \kappa^2 \left( \langle L_\alpha , L_\beta \rangle - \frac{1}{2} \eta_{\alpha\beta} \langle L_\gamma , \eta^{\gamma\delta} L_\delta \rangle   \right)  ,
  \end{equation} 
  and satisfies $\partial^\alpha T_{\alpha\beta} = 0$. In light-cone coordinates $\sigma^\pm = \tau \pm \sigma$ we have,
\begin{equation}
\begin{aligned}
T_{\pm\pm} = \kappa^2 \langle L_\pm , L_\pm \rangle, \qquad T_{+-} = 0 ,
\end{aligned}
\end{equation}
and,
\begin{equation} \label{eq:PCMEMconserved}
\partial_+ T_{--} = \partial_- T_{++} = 0 .
\end{equation}
Hence, classically, the theory is  conformal invariant.\footnote{Going to one-loop, however, the coupling $\kappa$ starts running and renders  the PCM for generic compact Lie groups asymptotically free \cite{Polyakov:1975rr,McKane:1979cm,Friedan:1980jf} as we have briefly mentioned at the end of the previous section. This is also clear from having a non-vanishing Ricci tensor. This is one of the reasons that the theory is considered  a toy model for QCD.}
Finally,  to work in the canonical formalism in the picture of  currents one has the Hamiltonian,
\begin{equation} \label{eq:PCMHamiltonian}
{\cal H}_{\text{\tiny PCM},\kappa^2} = - \frac{\kappa^2}{4\pi} \Big( \langle L_\tau , L_\tau \rangle + \langle L_\sigma , L_\sigma \rangle \Big) = - \frac{1}{2\pi} T_{\tau\tau},
\end{equation}
and  the fundamental equal-time Poisson brackets,
\begin{equation}\label{eq:PCMPoissonBracketsLeft}
\begin{aligned}
\left\{ L^A_\tau (\sigma) , L^B_\tau (\sigma ') \right\} &=   \frac{2\pi}{\kappa^2} F^{AB}{}_C L_\tau^C (\sigma) \delta(\sigma - \sigma ')  ,\\
\left\{ L^A_\tau (\sigma) , L^B_\sigma (\sigma ') \right\} &=  \frac{2\pi}{\kappa^2} \left(  F^{AB}{}_C L_\sigma^C (\sigma) \delta(\sigma - \sigma') -  \eta^{AB} \partial_\sigma \delta(\sigma - \sigma') \right) ,\\
\left\{ L^A_\sigma (\sigma) , L^B_\sigma (\sigma ') \right\} &=  0 ,
\end{aligned}
\end{equation}
where we expanded in the Lie algebra $\mathfrak{g}$.

    \begin{exercise}
Derive the Poisson brackets \eqref{eq:PCMPoissonBracketsLeft} using the canonical Poisson brackets defined through the canonical fields $x^\mu$ (the coordinates of target space) and momenta $\pi_\mu$. First introduce local coordinates $x^\mu$, $\mu \in \{ 1, \cdots , \dim G \}$ on the group manifold as $g = g(x^\mu(\sigma^\alpha))$ as above, interpret them as the canonical fields, and derive their canonical momenta $\pi_\mu = \frac{\partial L_{\text{\tiny PCM}, \kappa^2}(x^\mu , \partial_\alpha x^\mu )}{\partial ( \partial_\tau x^\mu))}$. Recall that we expand the Lie-algebra-valued worldsheet-one-forms as $L = -i L^A T_A = -i L_\alpha^A T_A d\sigma^\alpha$ and note that they can be pushed to target space as $L=  -i L_\mu^A T_A dx^\mu$ with $dx^\mu=\partial_\alpha x^\mu d\sigma^\alpha$. Show that,
\begin{equation}
\pi_\mu = \frac{\kappa^2}{2\pi} L_{\mu}^A \eta_{AB} L_\tau^B ,
\end{equation}
and therefore
\begin{equation}
L_\tau^A = \frac{2\pi}{\kappa} \eta^{AB} \pi_\mu L_B^\mu ,
\end{equation}
with $L_A^\mu$ the inverse of $L_\mu^A$, $L_\mu^A L^\mu_B = \delta^A_B$, and $\eta^{AB}$ the inverse of the bilinear form.
Recall that the canonical Poisson brackets are $\{\pi_\mu(\sigma) , x^\nu (\sigma')  \} = \delta(\sigma-\sigma') \delta_\mu^\nu$ and that for generic functions we have
\begin{equation}
\{ A(\sigma), B (\sigma') \} = \int d \sigma'' \frac{\delta A(\sigma)}{\delta \pi_\mu(\sigma'')} \frac{\delta B(\sigma')}{\delta x^\mu (\sigma'')} - \frac{\delta B(\sigma')}{\delta \pi_\mu(\sigma'')} \frac{\delta A(\sigma)}{\delta x^\mu(\sigma'')} ,
\end{equation}
which, as usual,  should be formally thought of as being integrated over $\sigma$ and $\sigma'$.
If you are stuck, take a look at this footnote.\footnote{Use the trick $f
(x) \partial_x\delta(x-y) = \partial_x \left( f(x) \delta(x-y) \right) - \partial_x f(x) \delta(x-y)$ and the Maurer-Cartan identity \eqref{eq:MCleft} expressed through $L_\mu^A$. Loosely speaking we also have $\partial_x \delta (x-y ) = -\partial_y \delta(y-x)$.}
    \end{exercise}   

\vspace{.2cm}

\noindent In terms of the right currents, for which the Maurer-Cartan identity \eqref{eq:MCright} has an extra minus sign in front of the commutator,\footnote{For this reason one can immediately infer their Poisson brackets by replacing $F_{AB}{}^C$ with $-F_{AB}{}^C$.} the Poisson brackets take the form,
\begin{equation}\label{eq:PCMPoissonBracketsRight}
\begin{aligned}
\left\{ R^A_\tau (\sigma) , R^B_\tau (\sigma ') \right\} &= - \frac{2\pi}{\kappa^2} F^{AB}{}_C R_\tau^C (\sigma) \delta(\sigma - \sigma ')  ,\\
\left\{ R^A_\tau (\sigma) , R^B_\sigma (\sigma ') \right\} &= - \frac{2\pi}{\kappa^2} \left(   F^{AB}{}_C R_\sigma^C (\sigma) \delta(\sigma - \sigma') +  \eta^{AB} \partial_\sigma \delta(\sigma - \sigma') \right) ,\\
\left\{ R^A_\sigma (\sigma) , R^B_\sigma (\sigma ') \right\} &=  0 .
\end{aligned}
\end{equation}
Notice that these Poisson brackets are non-ultralocal due to the presence of the term containing a derivative of the delta function.  At last, the Poisson algebra between the left and right currents is,\footnote{To derive these easily  (modulo tricks with the non-ultralocal term) one can use that $\{ L_\tau^A (\sigma) , g(\sigma' ) \} =   \frac{2\pi i}{\kappa^2} \eta^{AB} g (\sigma) T_B \delta(\sigma - \sigma')$, $\{ L_\tau^A (\sigma) , g^{-1}(\sigma' ) \} = -  \frac{2\pi i}{\kappa^2}  \eta^{AB}  T_B g^{-1}(\sigma)  \delta(\sigma - \sigma')$ and $R = g L g^{-1}$. See also \cite[Part II, Ch. 1, §5]{Faddeev:1987ph}. }
\begin{equation}\label{eq:PCMPoissonBracketsLeftRight}
\begin{aligned}
\left\{ L_\tau^A(\sigma) , R_\tau^B (\sigma' ) \right\} &= \left\{ L_\sigma^A (\sigma) , R_\sigma^B (\sigma' ) \right\} = 0 , \\
\left\{ L_\tau^A(\sigma) , R_\sigma (\sigma' ) \right\}  &= \frac{2\pi i}{\kappa^2} \eta^{AB} g T_B g^{-1} \partial_\sigma \delta (\sigma - \sigma' ) , \\
\left\{ L_\sigma^A(\sigma) , R_\tau (\sigma' ) \right\}  &= \frac{2\pi i}{\kappa^2} \eta^{AB} g T_B g^{-1} \partial_\sigma \delta (\sigma - \sigma' ) .
\end{aligned}
\end{equation} 
Notice that the Noether charges \eqref{eq:PCMfirstcharges} Poisson commute.

% In light-cone coordinates we have,
%\begin{equation}
%\begin{aligned}
%\delta S_{\text{\tiny PCM} , \kappa^2} &= \frac{\kappa^2}{\pi} \int_\Sigma d\sigma d\tau \langle g^{-1} \delta g ,  \partial_+ (g^{-1} \partial_- g ) + \partial_- (g^{-1} \partial_+ g ) \rangle, \\
%&= \frac{\kappa^2}{\pi} \int_\Sigma d\sigma d\tau \langle  \delta g g^{-1} ,  \partial_+ ( \partial_- g g^{-1} ) +   \partial_- ( \partial_+ g g^{-1} ) \rangle, 
%\end{aligned}
%\end{equation}

\subsection{Zero-curvature formulation: Lax connection and  towers of conserved charges}  \label{s:PCMint}
In the previous section we have seen in eq.~\eqref{eq:PCMEOM2} that the dynamics of the PCM model is captured by a flat and conserved current $L$. A classical integrable theory should, however, have a family of flat currents ${\cal L}(z)$ which control the dynamics for any value of a spectral parameter $z\in\mathbb{C}$ by
%As we have seen in the previous chapter, for classical integrability we need a reformulation of the dynamics in terms of a  Lax connection that depends on an arbitrary spectral parameter $z$ and satisfies a flatness condition,
\begin{equation} \label{eq:LaxFlat}
\mathrm{d} {\cal L}(z) + {\cal L}(z) \wedge {\cal L}(z) = 0, \qquad \forall z \in \mathbb{C} .
\end{equation}
The Maurer-Cartan identity for $L$ very much suggests that this will be possible for the PCM. Indeed, let us combine both equations of  \eqref{eq:PCMEOM2}   by considering the object
\begin{equation}
{\cal L}(\alpha, \beta ) = \alpha L + \beta \star L,
\end{equation}
where $\alpha, \beta \in \mathbb{C}$ are constants that we will fix such that the flatness condition \eqref{eq:LaxFlat} is satisfied upon the equations of motion. To have  integrability this system can not be fully determined: there should be at least one redundancy in the parameters  indicating the existence of a free (spectral) parameter. Using \eqref{eq:PCMEOM2} we find
\begin{equation}
\mathrm{d} {\cal L}(\alpha, \beta ) + {\cal L}(\alpha, \beta ) \wedge {\cal L}(\alpha, \beta ) = \left( \alpha^2 - \beta^2 -\alpha \right) L \wedge L ,
\end{equation}
which indeed can solve \eqref{eq:LaxFlat} with a single redundancy (we have one equation,  $\alpha^2 - \beta^2 -\alpha \overset{!}{=} 0$, for two variables). For $\alpha = \frac{1}{1-z^2}$ and $\beta = \frac{z}{1-z^2}$ we  find the Lax pair \cite{Pohlmeyer:1975nb,Zakharov:1973pp},
\begin{equation} \label{eq:PCMLax}
{\cal L}(z) = \frac{L + z \star L}{1- z^2},
\end{equation}
where the flatness condition \eqref{eq:LaxFlat} now implies the equations of motion \eqref{eq:PCMEOM2} for any value $z \in \mathbb{C}$.  Recall from section \ref{s:ClassicalIFT} that when $\Sigma = \mathbb{R}\times S^1$ the trace of  the monodromy matrix $T(2\pi, 0 ; z)$  build from the flat Lax connection (as in \eqref{eq:monodromy}) is,
\begin{equation}
T(2\pi, 0 ; z) = \overleftarrow{P \exp} \left( - \int^{2\pi}_{0} d\sigma \, \frac{L_\sigma + z L_\tau}{1-z^2} \right) ,
\end{equation}
 and generates an infinite tower of conserved charges because $\partial_\tau \Tr T(2\pi, 0 ;z)^n = 0$ for all $n\in\mathbb{N}_0$ and $z\in \mathbb{C}$. When on the other hand $\Sigma = \mathbb{R}^{1,1}$ the monodromy matrix,
\begin{equation}
T(+\infty, -\infty ; z) = \overleftarrow{P \exp} \left( - \int^{+\infty}_{-\infty} d\sigma \, \frac{L_\sigma + z L_\tau}{1-z^2} \right) ,
\end{equation}
 generates an infinite tower because  $\partial_\tau T(+\infty,-\infty ; z)^n = 0$ for all $n\in \mathbb{N}_0$ and $z \in \mathbb{C}$. By Taylor expanding  around appropriate values of the spectral parameter $z$ we can obtain various interesting sets of conserved charges, both local and non-local. We will illustrate this in what follows,  limiting ourselves to the  theory on the infinite line $\Sigma = \mathbb{R}^{1,1}$ and  expanding $T(z) \equiv T(+\infty, -\infty ; z)$.  Let us  remark  that the charges obtained in this way are not necessarily Poisson involutive. In that  case they generate a large hidden symmetry algebra, as we will briefly mention.\\
 
% \begin{equation}
%\begin{aligned}
%{\cal T}(z) &\equiv \Tr T(2\pi , 0 ;z )\\ 
%&= \Tr \left( 1 - \int^{2\pi}_0  d\sigma {\cal L}_\sigma (\tau, \sigma; z) + \int^{2\pi}_0 d \sigma \int^{\sigma'}_0 d\sigma' {\cal L}_\sigma (\tau, \sigma; z) {\cal L}_\sigma (\tau, \sigma' ; z) + \cdots \right) ,
%\end{aligned}
%\end{equation}

\noindent \textit{Remark.} Note that the Lax connection was easily constructed using  flat conserved currents, by simply combining $L$ and $\star L$. This is a  feature present in many well-known integrable deformations. For instance, in deformations of the PCM the current $L$ gets deformed, while it remains flat and conserved upon the equations of motion. The deformed Lax connection then takes a similar form as eq.~\eqref{eq:PCMLax} with $L$ replaced by its deformed version. See for example the recent lecture notes \cite{Hoare:2021dix} and some further comments and examples in section \ref{s:intdef}. Of course, note that this is not a necessary feature for an integrable field theory. Finally let us remark that when comparing with other references one can always redefine the free parameter $z$.  \\

\noindent  {\bf $\bullet$ Non-local charges: the Yangian ${\cal Y}_R (\mathfrak{g})$} --- By Taylor expanding $T(z)$ around $z = \infty$ we find a tower of non-local charges $Q_{(\infty)}^{(n)}$ defined by,
\begin{equation} \label{eq:YangianExpansion}
T(z) = 1+ \sum_{n = 0}^\infty \left( \frac{Q_{(\infty)}^{(n)}}{z^{n+1}} \right) .
\end{equation}
Explicitly we expand the  spatial component ${\cal L}_\sigma (z)$ of the Lax pair around $z = \infty$,
 \begin{equation}
 {\cal L}_\sigma (z) = - \frac{L_\tau}{z} - \frac{L_\sigma}{z^2} + {\cal O}(z^{-3}) ,
 \end{equation}
so that,
\begin{equation}
\begin{aligned}
T(z)={}&1 - \int^{+\infty}_{-\infty}  d\sigma {\cal L}_\sigma (\tau, \sigma; z) + \int^{+\infty}_{-\infty} d \sigma \int^{\sigma}_{-\infty} d\sigma' {\cal L}_\sigma (\tau, \sigma; z) {\cal L}_\sigma (\tau, \sigma' ; z) + \cdots \\
={}& 1 + \frac{1}{z} \int^{+\infty}_{-\infty} d\sigma \, L_\tau (\tau, \sigma) \\ &  + \frac{1}{z^2} \left( \int^{+\infty}_{-\infty} d\sigma \, L_\sigma (\tau, \sigma)   + \int^{+\infty}_{-\infty} d\sigma \int^{\sigma}_{-\infty} d\sigma'\, L_\tau (\tau, \sigma) L_\tau (\tau, \sigma')  \right) 
%\\
%&
 + {\cal O}(z^{-3}) \ , 
\end{aligned} 
\end{equation}
and hence,
\begin{equation}\label{eq:PCMnonlocalright}
\begin{aligned}
Q_{(\infty)}^{(0)} &= Q_R^{(0)} =  \int^{+\infty}_{-\infty} d\sigma \, L_\tau (\tau, \sigma), \\
Q_{(\infty)}^{(1)} &=  \int^{+\infty}_{-\infty} d\sigma \, L_\sigma (\tau, \sigma)   + \int^{+\infty}_{-\infty} d\sigma \int^{\sigma}_{-\infty} d\sigma'\, L_\tau (\tau, \sigma) L_\tau (\tau, \sigma') .
\end{aligned}
\end{equation}
Notice that the first charge corresponds precisely to the  conserved charge \eqref{eq:PCMfirstcharges} related to the global $G_R$ symmetry. The second  conserved charge is non-local: it is not a simple integral of a local density. An alternative but equivalent way to generate these---without using the monodromy matrix---is by a recursive relation for the related conserved currents as described in \cite{Brezin:1979am}. Let us discuss this method briefly.  Consider the  derivative $\nabla_\alpha = \partial_\alpha + L_\alpha $ which is flat, $[\nabla_\alpha , \nabla_\beta] = 0$, and covariant, $\partial^\alpha \nabla_\alpha \cdot = \nabla_\alpha \partial^\alpha \cdot$, by the equations of motion \eqref{eq:PCMEOM2}  of the PCM. Assuming we have the $n$th conserved current $L^{(n)}_\alpha$ of the integrable tower, $\mathrm{d}\star L^{(n)} = 0$, so that locally there exists a function $\chi^{(n)}$ satisfying,
\begin{equation} \label{eq:localsolconserv}
\star L^{(n)} =\mathrm{d} \chi^{(n)}, \qquad \epsilon^{\alpha}{}_{\beta} L^{(n)}_\alpha = \partial_\beta \chi^{(n)} ,
\end{equation}
 we can define the next conserved  current by,
\begin{equation}
L_\alpha^{(n+1)} = \nabla_\alpha \chi^{(n)},
\end{equation}
for $n\geq 0$. \newline

\begin{exercise}
Show that by setting  $L^{(0)}_\alpha \equiv L_\alpha$  one can prove by induction that $L_\alpha^{(n+1)}$  is indeed conserved as well.  Show that for $n=0$ and $n=1$ their associated charges are the same as the ones of \eqref{eq:PCMnonlocalright}     derived from the monodromy matrix.\footnote{Note that for $n=0$ eq.~\eqref{eq:localsolconserv} can be  solved by,
\begin{equation}
\chi^{(0)} = \int^\sigma d\sigma' L_\tau (\tau , \sigma' ) ,
\end{equation}
which is consistent upon the current conservation law, i.e.~$\partial_\sigma \chi^{(0)}  = L_\tau$ and $\partial_\tau \chi^{(0)}  = L_\sigma$ follows using $\partial_\tau L_\tau = \partial_\sigma L_\sigma$.} 
\end{exercise}

\vspace{.2cm}

 \noindent By  solving \eqref{eq:localsolconserv} for $\chi^{(n)}$ it is clear that, apart from the zeroth charge, all charges will be non-local. In addition, they are not necessarily Lie algebra valued as we see indeed for $Q^{(1)}_{(\infty)}$. This can be ``improved" by considering,\footnote{Here the bracket $\{ \cdot, \cdot \}$ is the anti-commutator and should not  be confused with the Poisson bracket.}
\begin{equation}
\begin{aligned}
\widetilde{Q}^{(1)}_{(\infty)} &= Q^{(1)}_{(\infty)} - \frac{1}{2} \int^{+\infty}_{-\infty} d\sigma \int^\sigma_{-\infty} d\sigma' \left\{L_\tau (\tau, \sigma) ,  L_\tau (\tau , \sigma')   \right\}, \\
&= \int^{+\infty}_{-\infty} d\sigma \, L_\sigma (\tau, \sigma)   + \frac{1}{2} \int^{+\infty}_{-\infty} d\sigma \int^{\sigma}_{-\infty} d\sigma'\, \left[ L_\tau (\tau, \sigma) ,  L_\tau (\tau, \sigma')  \right] .
\end{aligned}
\end{equation}
which is conserved as well. More systematically, the Lie algebra valued charges  can be constructed by redefining the covariant derivative as $\nabla_\alpha \cdot = \partial_\alpha \cdot + \left[ L_\alpha , \cdot \right]$ (which is flat as well)  \cite{MacKay:1992he,Evans:1999mj} and by starting the induction process by setting  first \textit{both} $L^{(0)}_{\alpha} \equiv L_\alpha$  and,
\begin{equation}
L^{(1)}_\alpha \equiv  \epsilon_{\alpha\beta} L^\beta + \frac{1}{2} \left[ L_\alpha , \int^{\sigma}_{-\infty} d\sigma' L_\tau (\tau , \sigma' ) \right] ,
\end{equation}
to prove that the currents $L_\alpha^{(n+1)} = \nabla_\alpha \chi^{(n)}$ are all conserved. 

The charges $Q_{(\infty)}^{(0)}$ and $\widetilde{Q}_{(\infty)}^{(1)}$ are also called the ``level-0" and ``level-1" charge because, interestingly, it was shown in \cite{MacKay:1992he}  that---under the Poisson brackets with fundamental relations given in \eqref{eq:PCMPoissonBracketsLeft}---they generate a classical Yangian algebra
 ${\cal Y}_R(\mathfrak{g})$.\footnote{At the quantum level, the algebra underlying the PCM and related to the $r$-matrix is a Yangian as well. The discussion done here can be understood as providing the classical origin of this Yangian algebra \cite{MacKay:1992he} .} {In particular, these non-local  charges are not Poisson involutive. At the quantum level they  are, on the other hand, vital to organise states into their representations and to constrain the S-matrix of the theory.}  Although this is a very interesting discussion, it is outside the scope of these lecture notes,
 % {\color{red}(apart from some brief comments that we give next),} 
 so we refer the unfamiliar reader to \cite{MacKay:2004tc,Loebbert:2016cdm} for a  pedagogical introduction. For comments on the Yangian algebra in the theory on $\Sigma = \mathbb{R}\times S^1$ see for instance \cite{Hatsuda:2004it,Hatsuda:2006ts}. \\
% In addition, notice that these non-local charges are not necessarily valued in the Lie algebra. With this is in mind one can  redefine the  covariant derivative as  \cite{MacKay:1992he},
%\begin{equation}
%\nabla_\alpha \cdot = \partial_\alpha \cdot +  \left[ L_\alpha , \cdot \right] ,
%\end{equation}
%which is still flat and covariant, leading  to the following  Lie algebra valued charges,
%\begin{equation}\label{eq:PCMnonlocalright2}
%\begin{aligned}
%Q_\infty^{(0)} &= Q_R^{(0)} =  \int^{+\infty}_{-\infty} d\sigma \, L_\tau (\tau, \sigma), \\
%Q_\infty^{(1)} &=  \int^{+\infty}_{-\infty} d\sigma \, L^{(1)}_\tau (\tau, \sigma),   \\ 
%&= \int^{+\infty}_{-\infty} d\sigma \, L_\sigma (\tau, \sigma)   + \int^{+\infty}_{-\infty} d\sigma \int^{\sigma}_{-\infty} d\sigma'\, \left[ L_\tau (\tau, \sigma) ,  L_\tau (\tau, \sigma')  \right] .
%\end{aligned}
%\end{equation}

%{\color{red}
%\noindent \textbf{Short Intermezzo: Yangian algebra.} The Yangian algebra of a Lie algebra is generated by level-0 $q^{(0)}$ and level-1 $q^{(1)}$ charges satisfying  \newline
%\begin{equation}
%\begin{aligned}
%\{ q_A^{(0)} , q_B^{(0)} \} &= i F_{AB}{}^C q_C^{(0)} ,  \\
%\{ q_A^{(0)} , q_B^{(1)} \} &= i F_{AB}{}^C q_C^{(1)} , 
%\end{aligned}
%\end{equation}
%}

\noindent  {\bf $\bullet$ Non-local charges: the Yangian ${\cal Y}_L (\mathfrak{g})$} --- Another interesting infinite set of non-local charges can be found  by considering the  gauge transformed Lax connection \eqref{eq:LaxGauge} and  monodromy matrix \eqref{eq:MonoGauge}. Taylor expanding $T^g(+\infty,-\infty  ; z) $ around $z = 0$  defines the set $Q_{(0)}^{(n)}$,
\begin{equation}
T^g(z)= 1 -   \sum_{n=0}^\infty z^{n+1} Q_{(0)}^{(n)}.
\end{equation}
Explicitly we have,
\begin{equation}
\begin{aligned}
{\cal L}_\sigma^g (z) &= g {\cal L}_\sigma(z) g^{-1} - \partial_\sigma g g^{-1} , \\ &= \frac{z^2 R_\sigma + z R_\tau}{1-z^2} , \\
&= z R_\tau + z^2 R_\sigma + {\cal O} (z^3),
\end{aligned}
\end{equation}
so that similarly as before the (Lie algebra valued) conserved charges are,
\begin{equation}
\begin{aligned}
Q_{(0)}^{(0)} &= Q_L^{(0)} = \int^{+\infty}_{-\infty} d\sigma R_\tau (\tau , \sigma), \\
\widetilde{Q}_{(0)}^{(1)} &=  \int^{+\infty}_{-\infty} d\sigma R_\sigma (\tau, \sigma) -  \frac{1}{2} \int^{+\infty}_{-\infty}  d\sigma  \int^{\sigma}_{-\infty} d\sigma' \left[ R_\tau (\tau , \sigma) , R_\tau (\tau , \sigma') \right] .
\end{aligned}
\end{equation}
The first conserved charge corresponds  to the global $G_L$ symmetry and the second conserved charge is  non-local. They generate again a classical Yangian algebra ${\cal Y}_L(\mathfrak{g})$ under the Poisson brackets which  commutes with the Yangian ${\cal Y}_R(\mathfrak{g})$ of above \cite{Wu:1982jt}. The global $G_L \times G_R$ symmetry of the PCM action can thus be thought of as sitting in a larger but hidden ${\cal Y}_L(\mathfrak{g}) \times {\cal Y}_R(\mathfrak{g})$ symmetry group. \\

\noindent \textit{Remark.} Note that, in the process of constructing the Classical Spectral Curve of the PCM, as briefly explained in section \ref{s:ClassMethods}, we can now derive the asymptotics of the quasimomenta easily. In the notation of footnote \ref{f:QMAsympt} we find for the PCM that
\begin{equation}
p_i(0) = 2\pi m_i , \qquad p_i'(0) = - Q_{L,i}^{(0)} , \qquad p_i(\infty) = Q_{R,i}^{(0)} \ ,
\end{equation}
with $m_i\in \mathbb{Z}$ and the components $Q_{L,i}^{(0)}$ and $Q_{R,i}^{(0)}$ the eigenvalues of $Q_{L}^{(0)}$ and $Q_{R}^{(0)}$ respectively.
Note that the $p_i(0)$ values follow from the intrinsic multi-valuedness in the definition of the quasimomenta, \eqref{eq:DefQM}.\\

\noindent {\bf $\bullet$ Local charges: higher spin and diagonal ones} --- An infinite tower of \textit{local} conserved charges can be obtained in the PCM  from several perspectives. Firstly, its classical conformality implies the conservation of powers of the energy and momentum since from \eqref{eq:PCMEMconserved} it follows that,
\begin{equation}\label{eq:SpinConservation}
\partial_\pm T_{\mp\mp}^p =0  ,
\end{equation}
for all $p \in \mathbb{N}$. These are associated to  higher-spin local charges as,
\begin{equation} \label{eq:LocalEMCharge}
Q^{(p)}_{(\pm)} = \int d\sigma \, T_{\pm\pm}^p , \qquad p \in \mathbb{N} ,
\end{equation}
which are conserved under both periodic and asymptotic boundary conditions. Secondly, the equations \eqref{eq:PCMEOM2}  for the flat conserved currents $L$ (or $R$)  are equivalent to
\begin{equation}
\partial_\pm L_\mp = \mp \frac12 [L_+ , L_- ] ,
\end{equation}
which implies that
\begin{equation} \label{eq:LocalLeftCurrentConservation}
\partial_\pm \Tr (L_\mp^s) = 0,
\end{equation}
for all $s\in \mathbb{N}$.\footnote{Note that for even $s$ eq.~\eqref{eq:LocalLeftCurrentConservation} coincides with \eqref{eq:SpinConservation}.} This provides an infinite set of local chiral conserved currents $\Tr (L_\pm^n)$ as discussed in \cite{Evans:1999mj,Evans:2000hx,Evans:2000qx}, and reflects again the fact that the theory is classically conformal invariant.
In fact for any rank-$s$ totally symmetric $\mathfrak{g}$-invariant tensor $d_{A_1 A_2 \cdots A_s}$ one has \cite{Evans:1999mj},
\begin{equation}\label{eq:EvansConservation}
\partial_\pm \left(  d_{A_1 A_2 \cdots A_s} L_\mp^{A_1} L_\mp^{A_2} \cdots L_\mp^{A_s} \right) = 0 .
\end{equation}
The spin of the associated charges is equal  to $s-1$.
Thirdly, and more generally, as discussed in section \ref{s:ClassicalIFT}, in any integrable field theory with a flat Lax connection local conserved chargers can be obtained by expanding the monodromy matrix $T(z)$ around poles of the Lax connection. Here the Lax \eqref{eq:PCMLax} has no regular part and the poles are located at $z = \pm 1$ with both of order $1$. In their vicinity one can perform a diagonal gauge transformation as in \eqref{eq:DiagonalGauge} to the diagonalised Lax connections ${\cal L}^{(\pm1)} (z)$. They can be expanded around $z = \pm1$ as,
\begin{equation}
{\cal L}^{(\pm1)} (z) = \sum_{n = -1}^\infty {\cal L }^{(\pm1)}_{n} (z \mp 1)^n .
\end{equation}
in which the coefficients correspond to conserved currents (see \eqref{eq:DiagZeroCurv}) implying the conservation of the  charges,
\begin{equation}
Q^{(\pm1)}_{n} = \int d\sigma \,   {\cal L }^{(\pm1)}_{n}{}_\sigma , \qquad n \geq -1 ,
\end{equation}
 which appear in the (diagonal) monodromy matrix as in \eqref{eq:DiagMono}. \\
\indent In the PCM there exists an additional structure within this set of charges. Notice that the light-cone components of the ordinary Lax \eqref{eq:PCMLax} are,
\begin{equation}\label{eq:PCMLaxLC}
\begin{aligned}
{\cal L }_\pm (z) &= \frac{L_\pm}{1\mp z} ,
\end{aligned}
\end{equation}
so that after diagonalising we have,
\begin{equation}
\begin{aligned}
{\cal L}_{-1}^{(\pm 1)}{}_\pm &= L^{(\pm 1)}{}_\pm , \qquad  &&{\cal L}_{n\geq 0}^{(\pm 1)}{}_\pm = 0 , \\
{\cal L}_{-1}^{(\pm 1)}{}_\mp &= 0 ,   &&{\cal L}_{n\geq 0}^{(\pm 1)}{}_\mp \neq 0 ,
\end{aligned}
\end{equation}
where $L^{(\pm1)}{}_\pm $ are the light-cone components of the  diagonalised left current around $z=\pm 1$. In particular $\mathrm{d} {\cal L}_{-1}^{(\pm1)}  = 0$ then implies \textit{chiral} conservation laws,
\begin{equation}
\partial_\mp {\cal L}_{-1}^{(\pm 1)}{}_\pm = \partial_\mp L^{(\pm1)}{}_\pm = 0 .
\end{equation}
They in turn give an alternative way to find the conservation laws given in \eqref{eq:LocalLeftCurrentConservation} with conserved charges,
\begin{equation}
Q^{(\pm 1)}{}^{(s)}  = \int d\sigma \, \Tr ( L_\pm^s ) ,
\end{equation}
in which we suppressed the $-1$ coefficient index. However, note that in contrast to the $s>1$ charges only $Q^{(\pm 1)}{}^{(s=1)}$ appears in the monodromy matrix as in \eqref{eq:DiagMono}. Conservation of energy and momentum, for instance, is contained in the higher spin local charges \eqref{eq:LocalEMCharge} for $s = 2p = 2$. The Hamiltonian \eqref{eq:PCMHamiltonian}  is proportional to,
\begin{equation}
{\cal H}_{\text{\tiny PCM},\kappa^2} \propto Q^{(+1)}{}^{(s=2)} + Q^{(-1)}{}^{(s=2)} ,
\end{equation}
and similarly the momentum is proportional to,
\begin{equation}
P_{\text{\tiny PCM},\kappa^2} \propto  Q^{(+1)}{}^{(s=2)} - Q^{(-1)}{}^{(s=2)} .
\end{equation}
How the higher spin charges appear precisely in the monodromy is not understood, while they traditionally are the ones important for quantum integrability. \\

\noindent \textit{Remark.} Again, in constructing the Classical Spectral Curve and the finite-gap equations, and in particular the resolvent,  one also needs the behaviour of the quasimomenta around the poles of the Lax connection. It is not difficult to derive that, for the PCM, the behaviour is
\begin{equation}
p_i (z) \approx \pm \frac{q_i^\pm}{1\mp z} , \qquad \text{around } \ z = \pm 1 \ ,
\end{equation}
with $q_i^\pm$ the components of the diagonal  matrix
\begin{equation}
q^\pm = \frac{i}{2} \int  d\sigma L^{(\pm1)}{}_\pm   \ .
\end{equation}
Because of \eqref{eq:LocalLeftCurrentConservation} we have $ L^{(\pm1)}{}_\pm =L^{(\pm1)}{}_\pm  (\tau \pm \sigma)$ and thus the  matrix $q^\pm$ is constant on the worldsheet.

\subsection {Involution of charges and the Maillet bracket}  \label{s:PCMMaillet}

 We have seen that  there are many different ways to construct an infinite set of conserved charges in the Principal Chiral Model, but to have (strong) classical integrability at least one the infinite sets should be in involution. As discussed in section \ref{s:FTInvolution}, it is sufficient that the 
 Poisson brackets of the Lax take a certain form (Sklyanin or Maillet). In that case the brackets between  the monodromy matrices $\{ \Tr_1 T_1(z) , \Tr_2 T_2(z')  \}$ vanishes and strong classical integrability is ensured. This is the direction that we will take, rather than showing Poisson involutivity of one of the above infinite sets of charges directly.\footnote{A direct examination of Poisson involutivity  of the higher spin local charges associated to \eqref{eq:EvansConservation} without relating to the monodromy matrix was done in \cite{Evans:1999mj}. It was found that these local charges do indeed give rise to an infinite commuting family when their spin equals the exponents of the Lie algebra modulo the Coxeter number. In addition it was shown in \cite{Evans:1999mj} that the local charges commute with the (Yangian) non-local charges so that the full classical symmetry is enlarged.}

 Let us proceed step by step by means of exercises. First, it will be convenient to rewrite the Poisson brackets of the left currents \eqref{eq:PCMPoissonBracketsLeft} in terms of the notation introduced at the end of section \ref{s:Liouville}. \newline

\begin{exercise}
Show that contracting \eqref{eq:PCMPoissonBracketsLeft} with $ T_A \otimes T_B$ gives,
\begin{equation}
\begin{aligned}
\{ L_{\tau, 1} (\sigma) , L_{\tau, 2} (\sigma') \} &=  \frac{2\pi}{\kappa^2} \left[ C_{12} , L_{\tau , 2} \right] \delta(\sigma - \sigma' ) , \\
\{ L_{\tau, 1} (\sigma) , L_{\sigma, 2} (\sigma') \} &=    \frac{2\pi}{\kappa^2} \left( \left[ C_{12} , L_{\sigma , 2} \right] \delta(\sigma - \sigma' )  + C_{12} \partial_\sigma \delta (\sigma - \sigma') \right)  ,\\
\{ L_{\sigma, 1} (\sigma) , L_{\sigma, 2} (\sigma') \} &= 0 .
\end{aligned}
\end{equation}
Note that $\{ L_{\sigma, 1} (\sigma) , L_{\tau, 2} (\sigma') \} =  \{ L_{\tau, 1} (\sigma) , L_{\sigma, 2} (\sigma') \}$.
This allows us to find the Poisson structure between the spatial component of the PCM Lax connection \eqref{eq:PCMLax} straightforwardly.
\end{exercise}

\vspace{.2cm}

\begin{exercise}
Show that,
\begin{equation} \label{eq:PoissonBracketsPCMLaxExpl}
\begin{aligned}
\{ {\cal L}_{\sigma, 1}(\sigma , z) , {\cal L}_{\sigma, 2}(\sigma' , z') \} ={}&   \frac{(2\pi/\kappa^2)}{(1-z^2) (1-z'^2)}  \Big( \left[ C_{12}, z z' L_{\tau, 2} + (z+z') L_{\sigma,2} \right] \delta(\sigma - \sigma')  \\
&  + (z+z') C_{12} \partial_\sigma \delta(\sigma - \sigma') \Big) ,
\end{aligned}
\end{equation}
which takes the form of the  $r/s$ Maillet bracket \eqref{eq:PoissonBracketsRSMaillet} with,
\begin{align}
r_{12}(z, z') &=   \frac{2\pi}{\kappa^2} \frac{C_{12}}{z-z'}  \phi^{-1}(z'),  \label{eq:PCMrmatrix}  \\
r_{21}(z', z) &=  \frac{2\pi}{\kappa^2} \frac{C_{12}}{z-z'}  \phi^{-1}(z) , \\
s_{12} (z,z') &= r_{12}(z, z') + r_{21}(z', z) =     -  \frac{2\pi}{\kappa^2} \frac{z+z'}{(1-z^2) (1-z'^{2})} C_{12} , \label{eq:PCMsmatrix}
\end{align}
and with $\phi(z) = \frac{1-z^2}{z^2} $ the so-called \textit{twist function}. As we have seen in exercise \ref{ex:rmatrix} the PCM $r$-matrix indeed satisfies the CYBE \eqref{eq:FTcYBE}. 
\end{exercise}

\vspace{.2cm}

\noindent From the work of Maillet  \cite{Maillet:1985ec,Maillet:1985ek} we are now  ensured the Principal Chiral Model is completely classically integrable in the strong sense.\\

 To end the discussion of the PCM, let us give a few small remarks on the quantum theory.
As mentioned in section \ref{s:FTInvolution}, the presence of the non-ultralocal term in \eqref{eq:PoissonBracketsPCMLaxExpl} highly   complicates the  application of the QISM which is traditionally used to quantise a theory whilst preserving integrability. Nevertheless, using the hidden Yangian algebra  together with unitarity and analyticity, one has been able to derive  the quantum $S$-matrix of the PCM  in \cite{Abdalla:1984iq,Wiegmann:1984ec},  and it was showed that it solves the QYBE. From a different perspective an interesting strategy to make progress has been developed by Faddeev  and  Reshetikhin in \cite{Faddeev:1985qu}. They ``regularise" or \textit{alleviate} the $r/s$ Maillet bracket so that the non-ultralocal term does not appear. In other words, one employs an alternative canonical structure which is not connected to the  PCM Lagrangian in the standard way. The alleviated theory, however,  can  be quantised without any additional subtle issues. It is then argued that the original classical PCM theory can be recovered by taking a non-trivial classical limit of the alleviated quantum theory (for more recent works, see also \cite{Delduc:2012qb}). Taking this non-trivial limit on the quantum results finally gives  quantum answers for the PCM. Alternatively, it might be that it is merely not suitable  to work with charges arising from the monodromy matrix. The local charges of \cite{Evans:1999mj} are possibly a good arena to pursue this  and circumvene  problems of non-ultralocality (see also \cite{Lacroix:2017isl}). Here the caveat is that one should analyse whether the conservation of the charges persists upon quantum effects, as initiated in \cite{Goldschmidt:1980wq} (see also \cite{Evans:1999mj}). For instance, the bosonic $CP^N$ sigma model does not remain integrable at the quantum level  for these reasons \cite{Goldschmidt:1980wq,Abdalla:1980jt}.

\subsection{A quick guide to integrable deformations} \label{s:intdef}

In recent years a lot of effort and progress has been made in the construction of a large family of  deformations of the Principal Chiral Model (and its generalisations\footnote{With generalisations of the Principal Chiral Model we are referring to the (semi-)symmetric space sigma models, in which a subgroup of the PCM's global symmetry group is gauged. These models are particularly important for realisations of string backgrounds relevant to the AdS/CFT correspondence. Pedagogical  introductions to these models can be found in e.g. \cite{Zarembo:2017muf,Hoare:LN}.}) which remain classically integrable for every value of the deformation parameter. A remarkable and interesting aspect in this line of research is the close connection with generalised worldsheet dualities (namely abelian, non-abelian and Poisson-Lie T-duality), see e.g.~\cite{Osten:2016dvf,Hoare:2016wsk,Borsato:2016pas,Vicedo:2015pna,Hoare:2015gda,Sfetsos:2015nya,Klimcik:2015gba,Hoare:2017ukq,Hoare:2018ebg}, and the possible reformulation of these theories within duality symmetric formulations of supergravity such as Double Field Theory \cite{Demulder:2018lmj,Borsato:2021gma,Borsato:2021vfy}. A strong motivation to study integrable deformations comes from the AdS/CFT correspondence, where one is encouraged by the fact that they can deform and break some of the unrealistic symmetries (conformal symmetries and supersymmetries) of well-known holographic  duals, whilst retaining the analytical (and, at the quantum level, exact) control provided by integrability. More formal motivations follow from  their close relation to fundamental mathematics 
 by providing Lagrangian formulations of theories possessing  infinite symmetries with intricate algebraic structures such as affine quantum groups, see e.g.~\cite{Delduc:2013fga,Delduc:2013qra,Hollowood:2015dpa}. 
 In this section we will give a very brief and quick guide to the most well-known integrable deformations. For much more details we refer to the recent lecture notes \cite{Hoare:2021dix} which, in addition, includes a more complete set of references to the recent literature. \\
 
 Most well-known  deformations have in common that they posses currents $A$ which are flat and conserved upon the equations of motion. As we have seen in section \ref{s:PCMint} this means these theories will possess a  Lax connection\footnote{Let us point out that the existence of a flat conserved current on-shell only implies (weak) classical integrability, and that this is also not a necessary condition. } 
 \begin{equation}
{\cal L}(z) = \frac{A + z \star A}{1- z^2}, \qquad
  {\cal L}_\pm(z) = \frac{A_\pm}{1\mp z} ,
 \end{equation}
 which is flat for every value of the spectral parameter. Prime examples are
 \begin{itemize}
 \item \textit{The Principal Chiral Model with Wess-Zumino term}. One can add a Wess-Zumino (WZ) term
 \begin{equation}
 + \frac{k}{12 \pi} \int_B d^3 y \epsilon^{\alpha\beta\gamma} \langle \bar{L}_\alpha , \bar{L}_\beta \bar{L}_\gamma \rangle ,
 \end{equation}
 to the PCM action \eqref{eq:actionPCML}. Here $\partial B = \Sigma$ and $\bar{L} = \bar{g}^{-1} d\bar{g}$ with the map $\bar{g}: B \rightarrow G$ an extension of $g:\Sigma \rightarrow G$ assuming that $\pi_2 (G)$  and $H_2 (G)$ are trivial, which assure the existence of the WZ term. Furthermore, we assume $H^3(G)$ integral, such that the level $k$ is quantised and  the path integral is insensitive to the choice of the extension $B$ \cite{Witten:1983ar}.  The flat conserved current is found to be \cite{osti}
 \begin{equation}
 A_\pm = \left(1 \mp \frac{k}{2\kappa^2} \right) L_\pm .
 \end{equation}
 Note that if we set $k=0$ the ``deformed" current $A_\pm$ reduces to the PCM current $L_\pm$. When, on the other hand, we set $\kappa^2 = k/2$ we obtain the exactly conformal Wess-Zumino-Witten (WZW) model  which has two chirally conserved currents generating two copies of a  Kac-Moody algebra. The latter is however subtle to see at the level of the currents $A_\pm$ because for $\kappa^2 = k/2$ we recover only the Kac-Moody current $L_-$. For more details on the WZW model see the original paper \cite{Witten:1983ar} but also chapter 15 of the book  \cite{DiFrancesco:1997nk}.
 \item \textit{The Yang-Baxter deformations}. A well-known integrable deformation is the (in)homogeneous Yang-Baxter deformation, which was originally constructed as an exemplar of a model exhibiting Poisson-Lie symmetry \cite{Klimcik:2002zj}. Only some years later it was realised that this model is also classically integrable \cite{Klimcik:2008eq}.  Its action principle takes the form
 \begin{equation}
 S = -\frac{\kappa^2}{4\pi} \int_\Sigma d^2\sigma \langle L_+, \left( 1 - \eta R_g \right)^{-1} L_- \rangle ,
 \end{equation}
 in which $\eta $ is the deformation parameter and $R_g = \text{Ad}_{g^{-1}} R \text{Ad}_g$ with $R:\mathfrak{g} \rightarrow\mathfrak{g}$ an antisymmetric operator which satisfies the modified Classical Yang-Baxter equation (mCYBE)
\begin{equation} \label{eq:YBmCYBE}
[Rx,Ry]-R([Rx,y]+[x,Ry])=-c^2 [x,y],\qquad\qquad
\forall x,y\in \mathfrak{g} \ .
 \end{equation} 
 Up to  rescalings of the  $R$-operator, $c^2$ can take the values $\{-1,0,1\}$. In the literature $c^2=1$ is referred to as the split case, while $c^{2} = -1$ is the non-split case. Both of them are referred to as the inhomogeneous Yang-Baxter deformations. When $c^2=0$, $R$ must satisfy the Classical Yang-Baxter equation\footnote{Equation \eqref{eq:YBmCYBE} for $c^2=0$ is equivalent to \eqref{eq:cYBE} but is written here in operator/matrix form, namely for $R:\mathfrak{g}\rightarrow \mathfrak{g}$. Note however that \eqref{eq:YBmCYBE} is not related to the Poisson bracket algebra of the model. } and the corresponding models are referred to as the homogeneous Yang-Baxter deformations \cite{Kawaguchi:2014qwa,vanTongeren:2015soa}\footnote{Homogeneous Yang-Baxter deformations are much richer in the sense that the CYBE allows more solutions for $R$ than the mCYBE, and all such solutions will correspond to a different deformed target space geometry.} which contain as a special case the well-known T-duality-Shift-T-duality (TST) transformations \cite{Osten:2016dvf}.
 The deformed current is \cite{Matsumoto:2015jja}
 \begin{equation}
 A_\pm = (1-c^2\eta^2) (1 \pm \eta R_g)^{-1} L_\pm .
 \end{equation}
 When turning off the deformation parameter $\eta$ we recover the PCM currents. This deformation has also been generalised  to (semi-)symmetric space sigma models which give rise to more general string backgrounds such as $AdS_5\times S^5$ in \cite{Delduc:2013fga,Delduc:2013qra}.
 \item \textit{The $\lambda$-deformation}. This deformation is an interpolation between the exactly conformal WZW model and the non-abelian T-dual of the Principal Chiral Model \cite{Sfetsos:2013wia}. Its action principle is
 \begin{equation}
 S = S_{WZW,k}  - \frac{k}{\pi} \int_\Sigma d^2 \sigma \langle R_+ , \left(\lambda^{-1} - \text{Ad}_{g^{-1} }\right)^{-1} L_- \rangle ,
 \end{equation}
where $S_{WZW,k}$ is the action of the WZW model. Its generalisation to (semi-)symmetric space sigma models was constructed in \cite{Hollowood:2014rla,Hollowood:2014qma}. Here $\lambda\in [0,1]$ is the deformation parameter which gives the WZW model for $\lambda \rightarrow 0$ and the non-abelian T-dual of the PCM for $\lambda \rightarrow 1$.    The deformed currents are \cite{Hollowood:2014rla}
\begin{equation}
A_\pm = \frac{2}{1+\lambda} \left( \lambda^{-1} - \text{Ad}_g^{\pm 1} \right)^{-1} g^{\pm 1} \partial_\pm g^{\mp} .
\end{equation}
Interestingly, the $\lambda$-deformation is related to the inhomogeneous Yang-Baxter deformation via Poisson-Lie T-duality \cite{Vicedo:2015pna,Hoare:2015gda,Sfetsos:2015nya,Klimcik:2015gba,Hoare:2017ukq,Hoare:2018ebg}.
For more details we refer to  the pedagogical lecture notes \cite{Thompson:2019ipl}.
 \item \ldots
 \end{itemize}

\noindent These theories share the property of a Maillet  structure for the Poisson brackets, in which the $r$-matrices all have the form of \eqref{eq:PCMrmatrix} but in which the twist function, and in particular the location of its poles, is deformed \cite{Maillet:1985ec,Vicedo:2010qd,Delduc:2012qb,Delduc:2012vq,Delduc:2014uaa,Appadu:2017xku}.\\

Exploring several approaches to classify  integrable models and finding an organising principle in the space of integrable theories, through e.g.~integrable deformations, is on its own a very active area of research (see e.g.~\cite{Vicedo:2017cge,Vicedo:2019dej,Lacroix:2020flf,deLeeuw:2021ufg,Borsato:2021vfy}). Interestingly many of the integrable deformations applied to  worldsheet sigma-models give rise to deformed backgrounds which satisfy the supergravity equations of motion. This opens several possibilities for the AdS/CFT correspondence, which are largely unexplored at the moment and which will hopefully provide exciting advancements in the future.

\newpage

\section{Concluding remarks} \label{s:Conclusions}

Classical integrability in two-dimensional field theories is a broad subject that we have introduced by starting from the definition of classical integrability in finite-dimensional classical systems, and which we later have illustrated with a particular two-dimensional sigma model, known as the Principal Chiral Model. In a finite-dimensional system we have formulated classical integrability in two (equivalent) ways: either through Liouville integrability, which requires as many (independent) conserved charges as degrees of freedom in involution, and through the Lax pair formulation, which requires a  particular representation of the equations of motion and the Poisson bracket structure. By taking the continuum limit,  and as such going to an integrable field theory, we have generalised the formulation of Lax pair integrability in terms of a zero-curvature formulation of the equations motion and showed how in this case one can construct an infinite tower of conserved charges. We have argued how the latter is already sufficient in order to apply classical integrable methods, such as the Classical Inverse Scattering Method and the Classical Spectral Curve, to solve the equations of motion or to obtain the semi-classical spectrum of the theory. To have strong integrability, one  requires  again a particular structure of the Poisson brackets, which prominently are either of Sklyanin or Maillet form. We introduced the Principal Chiral Model and  showed how  interesting algebraic structures can arise in the infinite tower of conserved charges. We finally ended with several comments on  prime examples of integrable deformations which may be relevant for the AdS/CFT correspondence. \\

There are many lines of research in the area of integrability that we have not (or barely) touched on in these lecture notes. The main directions that one can take are:
\begin{itemize}
\item To quantise finite-dimensional systems and thus  describe  the interaction of a finite number of particles. Usually one considers the particles to have a fixed spin and the interactions to take place in a compact space. In other words, one considers quantum spin chains with a finite number $N$ of sites. To define quantum integrability an important concept is that of locality, which only makes sense for $N\rightarrow \infty$. An integrable quantum spin chain is thus defined as one which possess an infinite number of local conserved quantities which all commute amongst each other. Again, there is a systematic way of constructing the tower of conserved quantities, namely by constructing a Lax and monodromy operator from an $r$-matrix solution. One can then apply the Quantum Inverse Scattering Method and the Algebraic Bethe ansatz, or the Coordinate Bethe ansatz, to find the energy spectrum and eigenvectors of the Hamiltonian, see e.g.~\cite{Faddeev:1996iy}. The prime example of an integrable quantum spin chain is the $\mathrm{XXX}_{1/2}$ Heisenberg spin chain, which appears in the $\mathfrak{su}(2)$ subsector of $N=4$ SYM with two scalar fields. This fact allowed to compute  two-point correlations functions in the planar limit easily (see e.g. the review \cite{Beisert:2010jr}).
\item To quantise  integrable field theories. As we have briefly mentioned in section \ref{s:qint}, the existence of an infinite tower of local higher spin charges at the quantum level leads to a factorised $2\rightarrow2$ $S$-matrix which does not allow the creation and annihilation of particles. There are several ways to find the $S$-matrix. One can construct it directly at finite loops by providing all scattering data of the particle types and use Feynman techniques, or use the Quantum Inverse Scattering Method. For integrable sigma models  this is usually difficult. Instead one   often uses the hidden non-local symmetries to bootstrap the S-matrix and check it against the QYBE and one-loop computations. Once obtained one can derive the spectrum of the quantum theories using methods   such as the Thermodynamic Bethe ansatz \cite{Yang:1968rm} and the Quantum Spectral Curve  \cite{Gromov:2013pga}. Both these methods account for periodic models or models with finite length. This is particularly important to match anomalous dimensions in AdS/CFT examples.  
\end{itemize}

I hope that these lecture notes provide a stepping stone  into the broad literature on integrability, and that they help to start  early-career researchers in resolving one of the many exciting open questions.  For comments, typos, and questions, you can get in contact with me at
 \href{mailto:sib.driezen@gmail.com}{sib.driezen@gmail.com}.

\section*{Acknowledgements}

I would like to thank the organising committee 
 of the XVII Modave Summer School in Mathematical Physics for inviting me to give these lectures, and the students for listening, their interesting questions, and their comments. It was a very welcoming experience to finally have another face-to-face teaching interaction.
 {\color{black} I would also like to thank Saskia Demulder and J.~Luis Miramontes for discussions, advice and  reading  the draft of these notes, and  Riccardo Borsato  for helpful  discussions.}
Parts of these notes are  based on my Ph.D.~thesis and thus I am also grateful to my Ph.D.~supervisors Daniel C.~Thompson and Alexander Sevrin.
I am supported by the fellowship of ``la Caixa Foundation'' (ID 100010434) with code LCF/BQ/PI19/11690019,
by AEI-Spain (under project PID2020-114157GB-I00 and Unidad de Excelencia Mar\'\i a de Maetzu MDM-2016-0692), by Xunta de Galicia-Conseller\'\i a de Educaci\'on (Centro singular de investigaci\'on  de  Galicia  accreditation  2019-2022, and project ED431C-2021/14), and  by the European Union FEDER.

\bibliographystyle{/Users/sibylledriezen/Dropbox/PhD:Bibfile/JHEP}
{ \bibliography{/Users/sibylledriezen/Dropbox/PhD:Bibfile/SibBib}
}

\end{document}